\documentclass[aps,pre,reprint,twocolumn,notitlepage,superscriptaddress]{revtex4-1}
\usepackage{epsfig,graphics,amssymb,amsmath,subeqnarray,color,upgreek}
\usepackage[colorlinks=true,linkcolor=blue]{hyperref}
\usepackage[normalem]{ulem}
\usepackage{soul}
\begin{document}

\title{Autophoretic locomotion in weakly viscoelastic fluids at finite P\'eclet number}
\author{Giovanniantonio Natale}
\affiliation{Department of Chemical and Petroleum Engineering, University of Calgary\\ Calgary, AB, T2N 1N4, Canada}
\affiliation{Department of Chemical and Biological Engineering, University of British Columbia\\ Vancouver, BC, V6T 1Z3, Canada}
\author{Charu Datt}
\affiliation{Department of Mechanical Engineering, University of British Columbia\\ Vancouver, BC, V6T 1Z4, Canada}
\author{Savvas G. Hatzikiriakos}
\affiliation{Department of Chemical and Biological Engineering, University of British Columbia\\ Vancouver, BC, V6T 1Z3, Canada}
\author{Gwynn J. Elfring}\email{gelfring@mech.ubc.ca}
\affiliation{Department of Mechanical Engineering, University of British Columbia\\ Vancouver, BC, V6T 1Z4, Canada}

\begin{abstract}
In this work, we numerically investigate the dynamics of a self-propelling autophoretic Janus particle in a weakly viscoelastic fluid. The self-propulsion is achieved by an asymmetry in the properties of the surface of the Janus particle that drives a surface slip velocity and bulk flow. Here we investigate the effect of viscoelasticity on this advection-diffusion problem over a range of P\'eclet and Damk\"ohler numbers. Particles are found to swim faster, or slower, in viscoelastic fluids, and we show how reaction and diffusion rates affect the viscoelastic stresses that lead to changes in propulsion.
\end{abstract}

\maketitle

\section{Introduction}
Active colloidal-scale particles that can locally convert chemical energy into motility thereby mimicking micro-organisms \citep{Def_AC} have received considerable attention in recent years for their potential applications as drug-delivery agents in biomedical applications  \citep{nelson2010microrobots}, as sensing and depolluting agents \citep{soler2013self, li2014water}, as microrheological probes \citep{behrend2004brownian} or as micromotors in micromachining systems  \citep{maggi2016self}. While many examples of active particles at microscales can be found in nature, such as bacteria like \emph{E. coli} and human spermatozoa, synthetic particles that can self-propel via phenomena like autophoresis or a self-induced Marangoni effect have also been created \citep{Stark_2016}.

Active particles need not depend on external field gradients as they themselves generate local gradients through chemical reaction (diffusiophoresis) or heat radiation (thermophoresis) on their surface \citep{anderson1989colloid, paxton2004catalytic, cordova2008osmotic,  ebbens2011direct, julicher2009comment, golestanian2005propulsion, golestanian2007designing}.  The field gradients can be produced using anisotropic properties on the particle surface \citep{golestanian2005propulsion, golestanian2007designing}. Experimentally, this anisotropy has been obtained by creating particles with surface compartments of different chemistries; such particles are popularly known as  Janus particles \citep{Janus_review}. The earliest studies of Janus particles focused on bimetallic (platinum-gold) rods which propel via electro-chemical surface reactions \citep{paxton2004catalytic, mei2011rolled}.  More recently, platinum half-coated spheres of silica or latex have also been synthesized \citep{howse2007self, ke2010motion, campbell2013gravitaxis}. In presence of hydrogen peroxide in a water environment, these particles are able to self-propel by creating an anisotropic distribution of solutes in their surroundings as metallic platinum (Pt) catalyses the decomposition of hydrogen peroxide. However, the exact mechanism of this propulsion -- electro or chemophoresis -- is still an open debate \citep{brown2014ionic, brown2015bulk}. If the platinum layer is substituted with a gold one, the particles can be heated up by exposure to UV light.  The propulsion is then due to locally induced temperature gradients. \citep{jiang2010active}.

Autophoretic propulsion has been well studied in Newtonian fluids\citep{Moran_2017}. A classical continuum approach was proposed by \citet{golestanian2007designing} wherein phoretic effects are accounted for as a distribution of slip velocities on the particle surface under the assumption that the interaction layer is thin compared to the particle size and advection of the solute is neglected. This framework was then extended to study advective effects \citep{cordova2008osmotic,julicher2009generic, cordova2013osmotic}, the role of geometry, \citep{popescu2010phoretic} and more complex surface reactions \citep{ebbens2012size}. For particles at the nanoscale the thickness of the interaction layer becomes comparable with the particle size and then careful considerations regarding the flow field, both within and without the boundary layer, need to be undertaken \citep{sabass2012dynamics, sharifi2013diffusiophoretic}. \citet{michelin14} recently analyzed the validity of the thin interaction layer assumption considering both advective and reactive effects. 

While much of the literature on autophoretic locomotion is limited to Newtonian fluids, active particles may also be found in complex polymeric fluids \citep{arratia_review}. In recent theoretical work, \citet{datt2017active} studied autophoretic locomotion in a weakly viscoelastic fluid. They found that a Janus particle can swim either slower or faster in a second-order fluid as compared to a Newtonian fluid depending on the distribution of surface activity. However, their analytical study neglected advection of the solute.  In another study, \citet{PRFStone} considered the motion of a hot Janus particle through a fluid with a spatially varying viscosity distribution due to a temperature difference between the particle and the ambient fluid and found that, in contradistinction to when the viscosity is uniform, the particle translational and rotational dynamics were coupled for the spherical Janus particle. Experimentally, \citet{gomez2016dynamics} studied silica spheres half-coated with carbon caps suspended in a binary mixture which displayed a finite relaxation time and shear-thinning behaviour. When illuminated, the fluid underwent local de-mixing causing autophoretic motion. They observed an increase of the rotational and translational diffusion coefficients with the increase in the particle velocity -- markedly different from the dynamics in Newtonian fluids. Apart from autophoretic particles, there have been several analytical and numerical studies on model microswimmers, see for example, recent reviews by \citet{elfring14} and \citet{sznitman2015}. Pertinent to this work are studies of squirmer model swimmers in complex fluids \citep{zhu2012self, montenegro, Li2014, decorato15, jfmCharu}, and in particular the study by  \citet{zhu2012self} who used the squirmer model\citep{Lighthill1951, Pedley_review} to understand the dynamics of puller and pusher-type swimmers in a viscoelastic (Giesekus) fluid. They showed that the viscoelastic swimming speed is lower than in Newtonian fluids owing to the presence of non-Newtonian extensional stresses. 

Autophoretic locomotion in viscoelastic fluids at finite P\'eclet number has not yet been investigated, and it is still unclear how the viscoelasticity of the fluid modifies the advection-diffusion in self-diffusiophoretic motion. Hence, the goal of this work is to understand the effect of viscoelasticity on the dynamics of chemically propelled particles at finite P\'eclet and Damk\"ohler numbers thereby extending analytical work by \citet{datt2017active}. To do so, we use a hybrid asymptotic-numerical method, combining a regular perturbation expansion to capture the leading-order effects of viscoelasticity while solving the Newtonian and weakly nonlinear viscoelastic concentration and flow fields numerically using a finite element method. We capture the effects of advection and surface activity on the propulsion velocity and provide a physical explanation for the impact of viscoelastic stresses.

\section{The diffusiophoresis problem}
In defining the diffusiophoretic problem, we closely follow the approach of \citet{michelin14}. Consider an isolated solid particle $\mathcal{B}$ with surface $\mathcal{\partial B}$, immersed in an otherwise quiescent fluid of viscosity $\eta$ and density $\rho$. The particle interacts with a solute dispersed in the fluid with concentration $C$, via short-range potential of characteristic range $\lambda$.  The solute is both advected by the fluid and diffuses with diffusivity $D$. We assume the chemical properties of the surface of the particle control the concentration flux by either a fixed-flux adsorption/desorption process with activity $A$ or a fixed-rate one-step chemical reaction with reaction rate $K$ \citep{michelin14}.

For a particle of size $a$ and density $\rho_s$ the Reynolds number $Re=\rho Ua/\eta$ and Stokes number $St=\rho_{s}Ua/\eta$ are assumed to be sufficiently small to neglect the inertia of both the fluid and solid. For the following discussion, the P\'eclet number $Pe$ represents the ratio of diffusive and advective time scales.

In this work we will assume a thin interaction layer limit $\epsilon = \lambda/a\ll 1$ for finite $Pe$. As shown by \citet{michelin14}, provided $\epsilon Pe\ll1$, advection within the interaction layer is negligible, and the solution of the advection-diffusion outside the interaction layer may be solved independently of the interaction layer dynamics by prescribing a slip velocity $\boldsymbol{u}^{s}$ at the solid particle boundary, given by
\begin{align}\label{Slipvelo}
\boldsymbol{u}^{s}=M \boldsymbol{\nabla}_s C,
\end{align}
where $\boldsymbol{\nabla}_s = (\boldsymbol{I}-\boldsymbol{nn})\cdot\boldsymbol{\nabla}$ is the projection of the operator onto the surface, and the local mobility,
\begin{align}
M=\pm\frac{k_{B}T\lambda^{2}}{\eta},
\end{align}
is defined from the local interaction potential profile, where $k_{B}$ is the Boltzmann constant and $T$ is the absolute temperature. The mobility can be either positive or negative depending on the form of the solute surface interaction with the particle; it is negative for locally attractive interactions, and positive for locally repulsive interactions \citep{michelin14}.

Under these conditions, the solute concentration $C$ is governed by the advection-diffusion equation and boundary conditions
\begin{align}
\frac{\partial C}{\partial t} + \boldsymbol{u} \cdot \nabla C &= D \nabla^{2} C,\label{conc}\\
D\boldsymbol{n}\cdot\boldsymbol{\nabla} C (\boldsymbol{x} \in \partial \mathcal{B})&=KC, \label{cocb}\\
C(r\longrightarrow \infty)&=C_{\infty},\label{conbc}
\end{align}
where $ C_{\infty}$ is the far field condition for the concentration. The distance from the centre of the particle $r = \left|\boldsymbol{x}-\boldsymbol{x}_0\right|$ where $\boldsymbol{x}_0$ is the center of the particle. Here we show a fixed-rate chemical reation at the surface but a fixed-flux process follows similarly as we shall show.

We assume the fluid is incompressible, and that inertia is negligible ($Re\ll 1$), therefore 
\begin{align}
\nabla \cdot \boldsymbol{u} &= 0,\label{continuity}\\
\nabla \cdot\boldsymbol{\sigma} &= \boldsymbol{0}.\label{velo}
\end{align}
The fluid is taken to be quiescent in the far field while the fluid velocity on the particle includes rigid-body translation $\boldsymbol{U}$ and rotation $\boldsymbol{\Omega}$ about $\boldsymbol{x}_0$,
\begin{align}
\boldsymbol{u}(r\longrightarrow \infty) &= \boldsymbol{0},\label{uc}\\
\boldsymbol{u}(\boldsymbol{x}\in\partial B) &= \boldsymbol{U}+\boldsymbol{\Omega}\times\boldsymbol{r}+\boldsymbol{u}^s,\label{ub}
\end{align}
where $\boldsymbol{r} = \boldsymbol{x}-\boldsymbol{x}_0$.

This set of equations is closed by noting that without inertia and absent any interparticle or external forces, the net hydrodynamic force and torque on the particle must be zero
\begin{align}
\boldsymbol{F} &= \int_{\partial\mathcal{B}}\boldsymbol{n}\cdot\boldsymbol{\sigma} dS=\boldsymbol{0},\\
\boldsymbol{L} &= \int_{\partial\mathcal{B}}\boldsymbol{r}\times(\boldsymbol{n}\cdot\boldsymbol{\sigma}) dS=\boldsymbol{0}.
\end{align}

It is convenient to recast the problem in dimensionless terms. To do so, we take $a$ as the characteristic length, while a natural scale for the concentration variations $a\mathcal{A}/D$, may be obtained from the flux boundary conditions. Here $\mathcal{A}$ is the characteristic magnitude of the surface activity. A velocity scale is obtained from the slip condition to be
\begin{align}
\mathcal{U} = \frac{\mathcal{A}}{D}\frac{k_{B}T\lambda^{2}}{\eta}.
\end{align}
From the velocity scale, we define a time scale $a/\mathcal{U}$ and stress scale $\eta \mathcal{U}/a$ (for both pressure and deviatoric stress). Introducing dimensionless variables, denoted by $^*$'s, we write
\begin{align}
c^{*}&=\frac{C-C_{\infty}}{\mathcal{A}a/D},\\
\boldsymbol{u}^* &= \boldsymbol{u}/\mathcal{U},\\
\boldsymbol{\sigma}^* &= \frac{\boldsymbol{\sigma}}{\eta \mathcal{U}/a},\\
t^* &= \frac{t}{a /\mathcal{U}}.
\end{align}

In dimensionless form, the governing equation for the concentration field and boundary conditions are
\begin{align}
Pe\left(\frac{\partial c^*}{\partial t^*} + \boldsymbol{u}^* \cdot \boldsymbol{\nabla}^* c^*\right) &= \nabla^{*2} c^*,\label{conc}\\
\boldsymbol{n}\cdot\boldsymbol{\nabla}^* c^* (\boldsymbol{x^*} \in \partial \mathcal{B})&=k\left(1+Da\, c^{*}\right),\label{cocbad}\\
c^{*}(r^{*}\longrightarrow \infty)&= 0,\label{conbcad}
\end{align}
where the dimensionless surface activity distribution $k=K/\mathcal{K}$, where $\mathcal{K}$ is a characteristic scale of the reaction, and thus we set $\mathcal{A}=\mathcal{K}C_\infty$. The P\'eclet number is defined as 
\begin{align}\label{Pe}
Pe=\frac{\mathcal{U}a}{D},
\end{align}
whereas the Damk\"ohler number
\begin{align}\label{Da}
Da = \frac{\mathcal{K}a}{D},
\end{align}
is the ratio between the diffusive and the reactive time scales. When $Da=0$ diffusion is fast enough that the far-field concentration sets the flux (fixed-flux); assuming a fixed-flux process at the outset we arrive at the same form except in that case $k=-A/\mathcal{A}$ (where $A$ is positive for adsorption) \citep{michelin14}.

The governing equations for the velocity field and boundary conditions in dimensionless form are
\begin{align}
\nabla \cdot\boldsymbol{\sigma}^{*}&=\boldsymbol{0},\label{veload}\\
\nabla \cdot \boldsymbol{u}^{*} &= 0,\label{continuityad}\\
\boldsymbol{u}^{*}(\boldsymbol{x^*} \in \partial \mathcal{B}) &= \boldsymbol{U}^{*}+\boldsymbol{\Omega}^{*}\times\boldsymbol{r}^{*}+\boldsymbol{u}^{s*}\label{ubad},\\
\boldsymbol{u}^{*}(r^{*}\longrightarrow \infty)&=\boldsymbol{0}.\label{ucad}
\end{align}

\section{Second-order fluid}
In this work we are interested in understanding the effects of viscoelasticity on advection-diffusion in the self-diffusiophoresis of a Janus particle. In order to gain insight into this complex problem, we will use the second-order fluid (SOF) constitutive model \citep{bird1977dynamics}, an asymptotic approximation of viscoelastic fluids for slowly varying flows. For a SOF, the stress is given by
\begin{align}\label{SO}
\boldsymbol{\sigma}=-p\boldsymbol{I}+\eta\dot{\boldsymbol{\gamma}}-\frac{\Psi_1}{2}\overset{\triangledown}{\dot{\boldsymbol{\gamma}}}+\Psi_2\dot{\boldsymbol{\gamma}}\cdot\dot{\boldsymbol{\gamma}},
\end{align}
where $\eta$ is the zero-shear viscosity, $\dot{\boldsymbol{\gamma}}$ is the strain-rate tensor, and $\Psi_{1}$ and $\Psi_{2}$ are the first and second normal stress-difference coefficients respectively. The upper-convected derivative
\begin{align}
\overset{\triangledown}{\dot{\boldsymbol{\gamma}}}=\frac{\partial\dot{\boldsymbol{\gamma}}}{\partial t}+\boldsymbol{u}\cdot\boldsymbol{\nabla}\dot{\boldsymbol{\gamma}}-(\boldsymbol{\nabla}\boldsymbol{u})^\top\cdot\dot{\boldsymbol{\gamma}}-\dot{\boldsymbol{\gamma}}\cdot\boldsymbol{\nabla}\boldsymbol{u}.
\end{align}
where $[\boldsymbol{\nabla}\boldsymbol{u}]_{ij}=\partial u_j/\partial x_i$.

In dimensionless form we obtain
\begin{align}\label{SOdim}
\boldsymbol{\sigma}^{*}=-p^*\boldsymbol{I}+\dot{\boldsymbol{\gamma}}^{*}-De\left(\overset{\triangledown}{\dot{\boldsymbol{\gamma}}^{*}}+b\dot{\boldsymbol{\gamma}}^{*2}\right),
\end{align}
where the Deborah number, $De=\Psi_{1}\mathcal{U} /2a\eta$, quantifies the departure from Newtonian behavior  \citep{decorato15, decorato16b} and $b=-2\Psi_{2}/\Psi_{1}$. The first normal stress coefficient, $\Psi_1$, is positive for nearly all polymeric fluids while $\Psi_2$ is typically negative and much smaller in magnitude \cite{bird95}; here we take $b=0.2$ to match the work by \citet{decorato15}. Finally, we note that while the second-order fluid model model represents a common form for most viscoelastic fluids for small Deborah number, it is expected to be valid only for very small strain-rates \citep{bird1977dynamics,leal80}.

Henceforth, we work in dimensionless quantities and drop the stars $^*$'s for convenience.

\section{Squirmer model}
We consider here spherical, axisymmetric Janus particles characterized by a surface activity disbribution $k=k(\mu)$  where $\mu=\cos \theta$ and $\theta$ is the polar angle with respect to the axis of symmetry $\boldsymbol{e}_{z}$ in spherical coordinates. For simplicity, we consider one side of the Janus particle to be reactive, $\mu \ge \mu_c$ whereas the other side is inert. The solute concentration and velocity field are hence also axisymmetric, $c=c(r,\mu)$ and $\boldsymbol{u} = u_r(r,\mu)\boldsymbol{e}_r+u_\theta(r,\mu)\boldsymbol{e}_\theta$, as is the tangential slip $\boldsymbol{u}^s = u_\theta(1,\mu)\boldsymbol{e}_\theta$. The active particle therefore undergoes translational motion $\boldsymbol{U} = U\boldsymbol{e}_z$ without rotation $\boldsymbol{\Omega}=\boldsymbol{0}$.

It is common to represent the tangential slip $u_{\theta}$ in terms of Legendre polynomials, or squirming modes
\begin{align}\label{alpha}
\alpha_{n}=\frac{1}{2}\int^{1}_{-1}\sqrt{1-\mu^{2}}L'_{n}(\mu)u_{\theta}(1,\mu)d\mu.
\end{align}

When $De=Pe=Da=0$, the flow is linear, the concentration field is harmonic, and an analytical solution is easily obtained that yields the swimming velocity
\begin{align}\label{cPe0}
			\boldsymbol{U}=\frac{M}{4}\left(1-\mu_{c}^{2}\right)\boldsymbol{e}_{z}, 
\end{align} 
where $M=\pm1$ in dimensionless form. This result may be found simply by way of the reciprocal theorem\citep{michelin14}, but outside this limit numerical solution is required.

\section{Reciprocal theorem}
The reciprocal theorem of low Reynolds number hydrodynamics \citep{happel65} may be used to find the rigid-body motion of a particle given a slip velocity $\boldsymbol{u}^s$ \citep{stone96,lauga09,lauga14,elfring16}. Following \citet{datt2017active}, we know that for a spherical particle of radius $a$ the translational velocity in a weakly nonlinear viscoelastic fluid is given by
\begin{align}
\boldsymbol{U} = &-\frac{1}{4\pi }\int_{\partial\mathcal{B}}\boldsymbol{u}^s d S - \frac{De}{8\pi}\int_{\mathcal{V}}\boldsymbol{\tau}_{NN} :\left(1+\frac{1}{6}\nabla^2\right)\boldsymbol{\nabla}\boldsymbol{G} d V
\end{align}
where $\boldsymbol{G} = (\boldsymbol{I}+\boldsymbol{r}\boldsymbol{r}/r^2)/r$ while the tensor $\boldsymbol{\tau}_{NN} = -\left(\overset{\triangledown}{\dot{\boldsymbol{\gamma}}^{*}}+b\dot{\boldsymbol{\gamma}}^{*2}\right)$ represents the weakly nonlinear contribution to the stress \cite{elfring17}. For a Newtonian fluid, $De=0$, the translational velocity is given simply by (minus) the surface average of the given slip velocity \cite{stone96,elfring15}, in which case only the first squirming mode contributes to the swimming speed
\begin{align}
\boldsymbol{U} = \alpha_1 \boldsymbol{e}_z.
\end{align}
When the fluid is non-Newtonian, integration of the tensor $\boldsymbol{\tau}_{NN}$ over the entire fluid domain is required, and in general all squirming modes may affect $\boldsymbol{U}$. As shown by \citet{datt2017active}, the leading-order correction to the swimming speed for a squirmer in a second-order fluid is given by
\begin{align}
\boldsymbol{U} = \alpha_1 \boldsymbol{e}_z +De(b-1)\sum_{p=1}^\infty C_p \alpha_p\alpha_{p+1} \boldsymbol{e}_z
\end{align}
where $C_p = 6p/[(p+1)^2(p+2)]$.

\section{Perturbative analysis}
In order to evaluate the leading-order effects of viscoelasticity on the advection-diffusion process in self-diffusiophoresis, we focus only on weakly nonlinear effects for small Deborah numbers and study the problem perburbatively about the Newtonian limit. We write the unknown velocity, pressure and concentration fields and the velocity of the particle as regular expansions in $De$ number,
\begin{align}
\boldsymbol{u}&=\boldsymbol{u}_{0}+De\,\boldsymbol{u}_{1}+O\left(De^{2}\right),\label{uexp}\\
p&=p_{0}+De\,p_{1}+O\left(De^{2}\right),\label{pexp}\\
c&=c_{0}+De\,c_{1}+O\left(De^{2}\right),\label{cexp}\\
U&= U_0+De\,U_1++O\left(De^{2}\right).
\end{align}

In this work we will consider all fields steady when moving with the velocity of the swimmer, for example the concentration field
\begin{align}
\frac{\partial c}{\partial t}+\boldsymbol{U}\cdot\boldsymbol{\nabla} c = \boldsymbol{0}.\label{steady}
\end{align}

\subsection{Zeroth order (a Newtonian fluid)}

At zeroth order, we have the diffusiophoretic motion of a Janus particle in a Newtonian fluid governed by the following set of equations
\begin{align}
\nabla^{2}\boldsymbol{u}_{0}&=\boldsymbol{\nabla} p_{0},\label{veloN}\\
\boldsymbol{\nabla} \cdot \boldsymbol{u}_{0} &= 0\\
Pe\,\boldsymbol{u}_{0}' \cdot \boldsymbol{\nabla} c_{0} &= \nabla^{2} c_{0},\label{concN}
\end{align}
where we have used the fact that the fields are steady in the moving frame \eqref{steady} and defined $\boldsymbol{u}_{0}' = \boldsymbol{u}_{0}-\boldsymbol{U}_{0}$. The boundary conditions, to leading order, are 
\begin{align}
c_{0}\left(r\longrightarrow \infty\right)&= 0,\label{conbcN}\\
\boldsymbol{u}_{0}\left(r\longrightarrow \infty\right)&= \boldsymbol{0},\\
	\frac{\partial c_{0}}{\partial r}(r=1)&=k(1+Da\,c_{0}), \label{cocbN}\\
	\boldsymbol{u}_{0}(r=1)&=\boldsymbol{U}_{0}+M\boldsymbol{\nabla}_s c_{0}.
\end{align}
These equations are closed by the the force-free condition on the particle, which, by the reciprocal theorem, equates to
\begin{align}
\boldsymbol{U}_{0}=-\left\langle\boldsymbol{u}_0^{s}\right\rangle = -M\left\langle \boldsymbol{\nabla}_s c_{0} \right\rangle,
\end{align}
where the angle brackets $\left\langle \cdot\right\rangle$ represent the surface average on the particle. 

\subsection{First order}
The governing equations for the first-order viscoelastic pertubation fields are
\begin{align}
-\boldsymbol{\nabla} p_1 +\nabla^2\boldsymbol{u}_1 &= \boldsymbol{\nabla} \cdot \left(\overset{\Delta}{\dot{\boldsymbol{\gamma}}}_{0}+b\dot{\boldsymbol{\gamma}}^{2}_{0}\right),\label{velSof}\\
\boldsymbol{\nabla} \cdot \boldsymbol{u}_{1} &= 0,\label{massSof}\\
Pe\left(\boldsymbol{u}_{0}' \cdot \boldsymbol{\nabla} c_{1}+\boldsymbol{u}_{1}' \cdot \boldsymbol{\nabla} c_{0}\right) &= \nabla^{2} c_{1},\label{cSof}
\end{align}
where $\boldsymbol{u}_{1}'=\boldsymbol{u}_{1}-\boldsymbol{U}_{1}$. The boundary conditions for the first-order problem are
\begin{align}
c_{1}(r\longrightarrow \infty)&=0,\\
\boldsymbol{u}_{1}(r\longrightarrow \infty) &= \boldsymbol{0},\\
\frac{\partial c_{1}}{\partial r}(r=1) &= k\,Da\,c_{1},\label{conbcSof}\\
\boldsymbol{u}_{1}(r=1) &= \boldsymbol{U}_1+M\boldsymbol{\nabla}_s c_{1}.\label{cocbSof}
\end{align}
These equations are closed by enforcing the force-free condition at first order in Deborah number, which may be restated by the reciprocal theorem as a condition on the translational velocity
\begin{align}\label{Usof}
\boldsymbol{U}_{1}&=-\left\langle\boldsymbol{u}_1^{s}\right\rangle - \frac{1}{8\pi}\int_{\mathcal{V}}\boldsymbol{\tau}_{NN}[\boldsymbol{u}_0] :\left(1+\frac{1}{6}\nabla^2\right)\boldsymbol{\nabla}\boldsymbol{G} d V
\end{align}
where $\left\langle\boldsymbol{u}_1^{s}\right\rangle = M\left\langle \boldsymbol{\nabla}_s c_{1} \right\rangle$. Here we can see that there are two effects due to viscoelasticity which can cause a change in the translational velocity of the particle. The first term on the right-hand side represents the change in the slip velocity of the particle due to the viscoelasticity of the fluid, while the second term represents the change due to viscoelastic stresses. It is important to stipulate here that we assume that the mobility remains constant, in other words we do not account for the impact of viscoelasticity on the boundary layer problem that governs the slip, and we restrict our attention only to changes in the slip velocity due to a viscoelastic perturbation of the concentration field $c_1$. Likewise, we assume that the solute diffusivity $D$ remains constant throughout the flow.

Note that in the case of $Da=0$, there is no concentration flux at the surface of the particle for the first-order concentration field. Moreover, when $Pe=Da=0$ there is no first-order change of the concentration field, and consequently no change in the slip velocity.

\section{Numerical method}

The zeroth and first-order problems are both solved numerically by employing a finite element method (FEM) technique. FEM calculations are performed using Mathematica software.  A Taylor-Hood (P2-P1) triangular 2D mesh in polar coordinates was generated to solve both the zeroth and first-order problem. By generating the mesh in polar coordinates, the elements of the mesh are automatically finer at the particle surface and wider at large $r$. The mesh refinement was verified to eliminate any dependence of our results on the mesh element size. For all simulations, we meshed a circular 2D domain around our colloidal particle of the size $20a$, discretized by approximately $10^4$ triangular elements.

We first solve the zeroth-order problem and then use the zeroth-order fields as a known input to the first-order problem. We also use an iterative scheme whereby the translational velocity given by the exact solution when $Pe=Da=0$ is used as an initial value, the advection-diffusion problem is solved and then the translational velocity is updated to satisfy the force-free condition. At each iteration, the velocity and concentration field are calculated. The solution is assumed to have converged when difference in the velocity is less than $10^{-6}$ (note that the problem is non-dimensionlized such that the magnitude of the velocity is expected to be $O(1)$).

For validation we ensure that our results match those of \citet{michelin14} for Newtonian fluids, and that our numerical results match the analytical results for viscoelastic fluids when $Pe=Da=0$ in \citet{datt2017active}. The numerical method presented above is then employed to explore for the first time the effects of viscoelasticity on the solute concentration and flow field around Janus particles for nonzero values of $Pe$ and $Da$, and ultimately to determine the effect on particle propulsion.

\begin{figure}[t!]
\centering
        \includegraphics[width=0.45\textwidth]{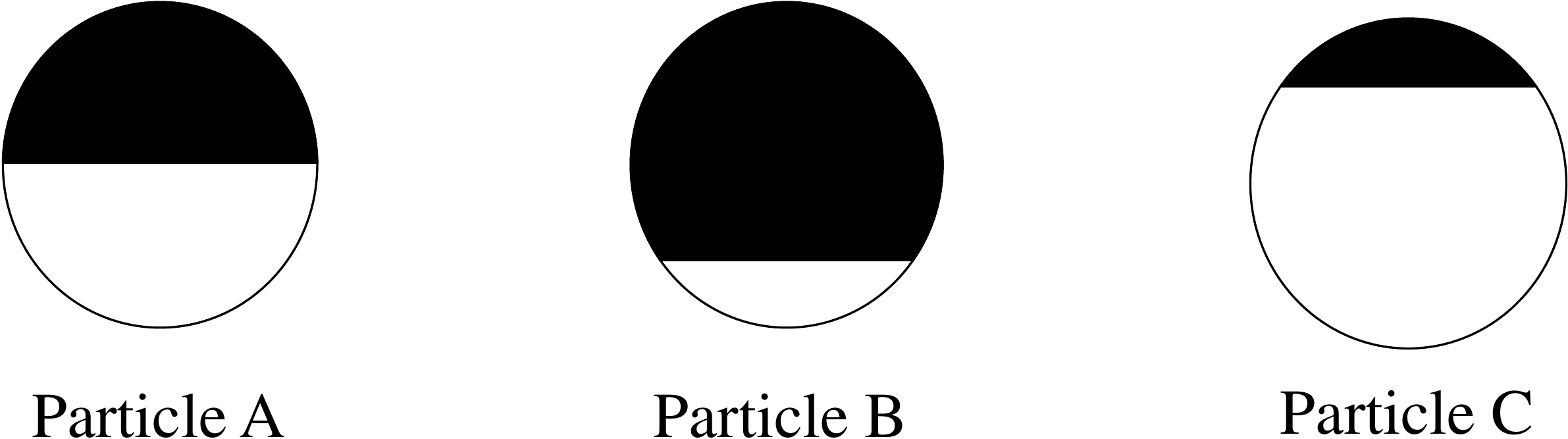}
        \caption{Schematic of the three Janus particles investigated. Black indicates the active region, while white is inert.}
		\label{JanusABC}
\end{figure}

\section{Results}
\begin{figure*}[t!]
\centering
        \includegraphics[width=\textwidth]{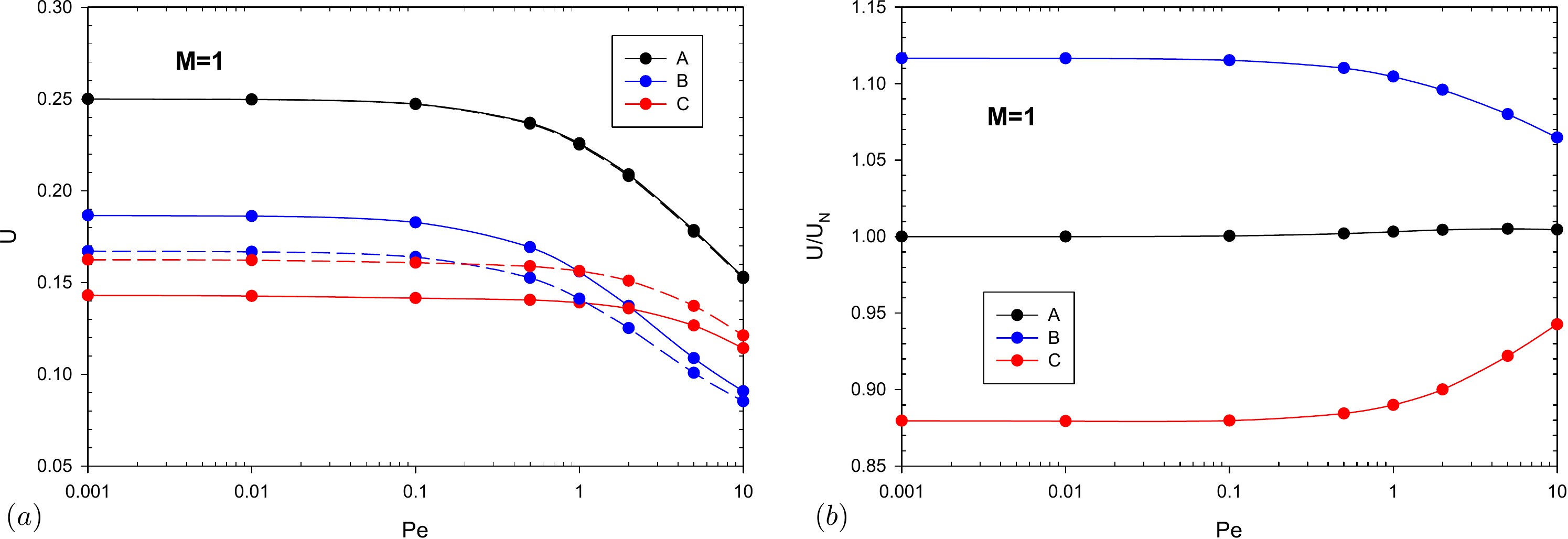}
        \caption{a) Propulsion speed as function of Pe for particle A, B and C in the case of $M=1$. Dashed and solid lines represent the propulsion in a Newtonian fluid, and a viscoelastic fluid ($De=1$), respectively. b) Ratio of viscoelastic and Newtonian propulsion velocity as function of $Pe$ for particle A, B and C in the case of $M=1$, with $De=1$.}
        \label{speedpos}
\end{figure*}

\begin{figure*}[t!]
\centering
        \includegraphics[width=\textwidth]{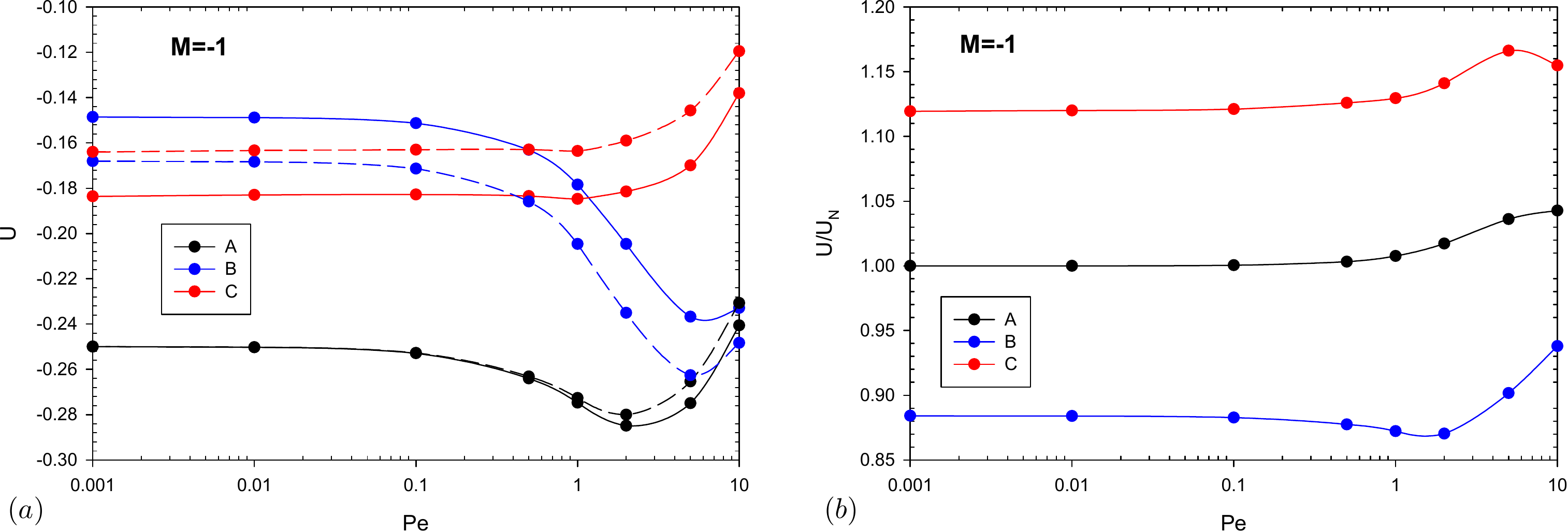}
        \caption{a) Propulsion speed as function of Pe for particle A, B and C in the case of $M=-1$. Dashed and solid lines represent the propulsion in a Newtonian fluid, and a viscoelastic fluid ($De=1$), respectively. b) Ratio of viscoelastic and Newtonian propulsion velocity as function of $Pe$ for particle A, B and C in the case of $M=-1$, with $De=1$.}\label{speedneg}
\end{figure*}

The Janus particles investigated here consist of an active cap while the rest of the particle is inert. In particular, three cases are considered: particle A is a symmetric Janus particle $\left(\mu_{c}=0\right)$, while particle B and C are asymmetric with $\mu_{c}$ equal to $-1/\sqrt{3}$ and $1/\sqrt{3}$, respectively (see Figure \ref{JanusABC}). Among the three Janus particles, particle B has the least inert surface coverage while particle C has the least active coverage. For all the particles, the mobility is considered uniform $\left(M = \pm 1\right)$. Thus, six different configurations are investigated in total.

In this work, the solute is always consumed by chemical reaction at the reactive cap; thus, for particles with positive mobility this implies that the slip velocity is oriented from the active cap to the inert surface. Consequently, the particles propel in the direction of the reactive cap ($\boldsymbol{e}_z$, the active end is the front). Particles with negative mobility move in the direction of the inert end ($-\boldsymbol{e}_z$, the inert end is the front). 

In the following, the results are divided in two main sections. First, we analyze the dynamics of the active particles with fixed-flux surface activity $\left(Da=0\right)$ and then probe the effect of nonzero $Da$ at low $Pe$. For clarity, in this paper we take $De=1$ in all figures so that the effects of viscoelasticity are more visually apparent, but note the results from our perturbative approach are strictly valid only when $De\ll 1$. Furthermore, when we refer to the velocity in a viscoelastic fluid (for example), we mean the velocity to first order in Deborah number $\boldsymbol{U}=\boldsymbol{U}_0+De\,\boldsymbol{U}_1$ since we neglect higher-order terms whose effects may be significant when the Deborah number is not very small \citep{decorato15}.

\subsection{Fixed-flux surface activity ($Da=0$)}

The effect of advection on autophoretic locomotion for fixed-flux surface activity $\left(Da=0\right)$ is analyzed here. The propulsion speed, in the $\boldsymbol{e}_z$ direction, of the three Janus particles immersed in a Newtonian and a second-order fluid is shown for positive mobility in Figure \ref{speedpos}a and negative mobility in Figure \ref{speedneg}a, while the ratio of viscoelastic to Newtonian propulsion speed is shown for positive mobility in Figure \ref{speedpos}b, and negative mobility in Figure \ref{speedneg}b. Positive values indicate propulsion towards the active pole ($M=1$), while negative values indicate propulsion towards the inert pole ($M=-1$).

For each of the three particles with positive mobility (particles swimming towards their active pole), the swimming speed in both Newtonian and viscoelastic fluid decreases monotonically with increasing $Pe$ number and at high $Pe$, a theoretical scaling of $U \propto Pe^{-1/3}$ is recovered in a Newtonian fluid \citep{julicher2009generic, michelin14,yariv15}. In contrast, with negative mobility (particles swiming towards their inert pole) the swimming speed displays a non-monotonic variation with $Pe$, with a maximum when $Pe = O(1)$ before the large P\'eclet asymptotic decay, $U \propto Pe^{-1/3}$.
    
\begin{figure}[t!]
\centering
        \includegraphics[width=.5\textwidth]{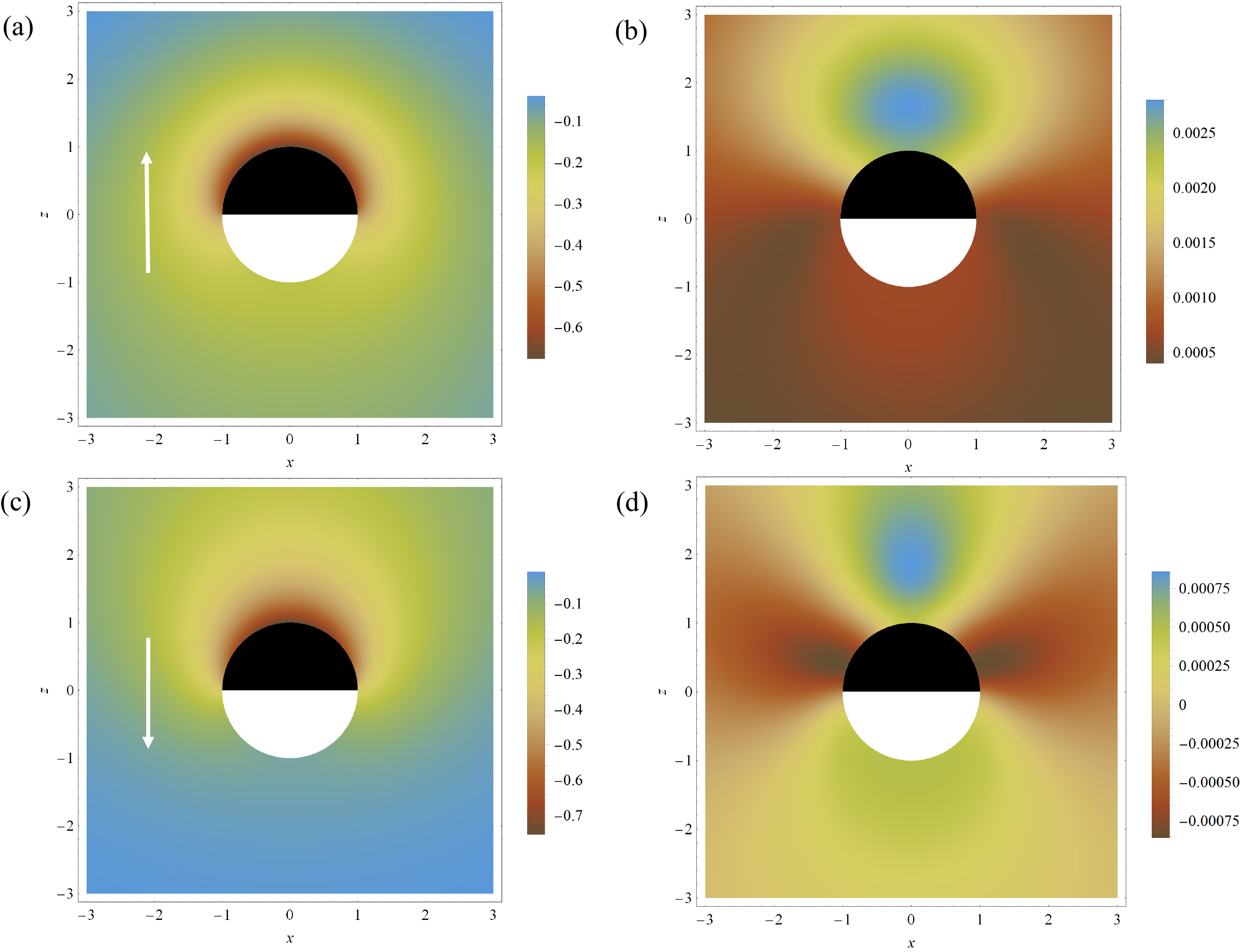}
        \caption{Relative solute concentration distribution around particle A in the Newtonian case (a,c) and second-order fluid contribution (b,d) for $Pe=2$ and $Da=0$. (a,b) represent the case of $M=1$ while (c,d) show the results for $M=-1$. The white vector indicates the direction of motion.}\label{ConcA}
\end{figure}

The effect of viscoelasticity on the translational velocity of the particle clearly depends on the coverage of the reactive cap. As shown in \citet{datt2017active}, a symmetric particle (Particle A) sees no viscoelastic effect on the velocity for $Pe=0$, and here we find a negligible change for nonzero P\'eclet numbers as well, when the mobility is positive. When the mobility is negative, we see a (relatively weak) viscoelastic effect at higher P\'eclet numbers. For asymmetric particles, viscoelasticity indeed makes a difference, particle B sees a speed increase (decrease) with respect to the Newtonian case while particle C, a speed decrease (increase), when the particle has positive (negative) mobility. For particles with positive mobility, the largest difference between the Newtonian and viscoelastic results occur when $Pe=0$, where our results match the analytical results in \citet{datt2017active}, and the difference in velocity due to viscoelasticity, strictly diminishes with increasing P\'eclet number. For particles with negative mobility, the effect of the advection of the solute and viscoelasticity is less straightforward with a non-monotonic change in the ratio of viscoelastic to Newtonian swimming speeds with increasing P\'eclet numbers.

\begin{figure*}[t!]
\centering
		\begin{minipage}[t]{0.48\textwidth}
        \includegraphics[width=\textwidth]{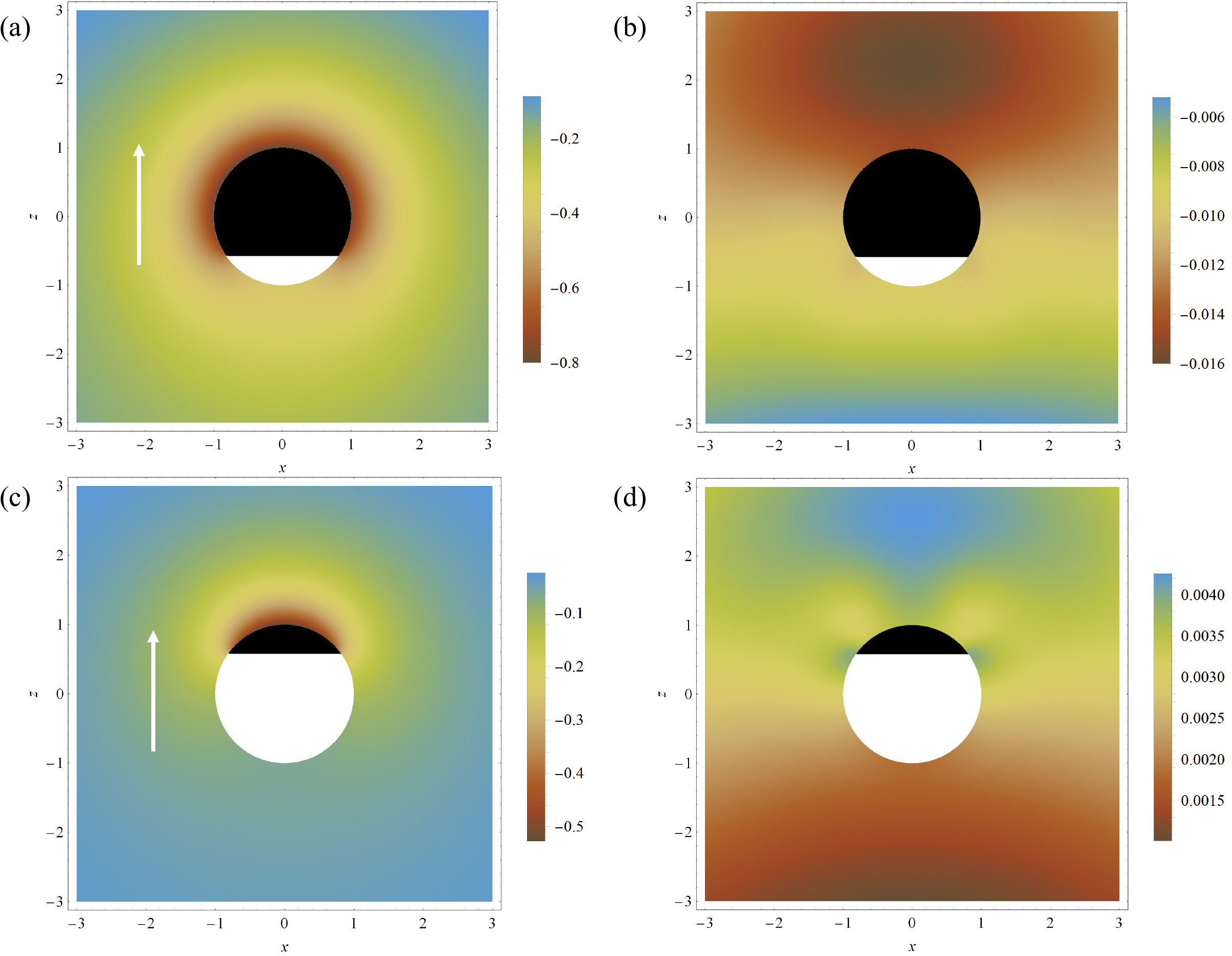}
        \caption{Relative solute concentration distribution around particle B and C in the Newtonian case (a) and (c) and second-order fluid contribution (b) and (d), respectively. The results are obtained for $Pe=2$ and $Da=0$ in the case of $M=1$. The white arrow indicates the direction of motion.}\label{ConcMp1}
        \end{minipage}
    \hspace{10pt}
	\begin{minipage}[t]{0.48\textwidth}
	    \includegraphics[width=\textwidth]{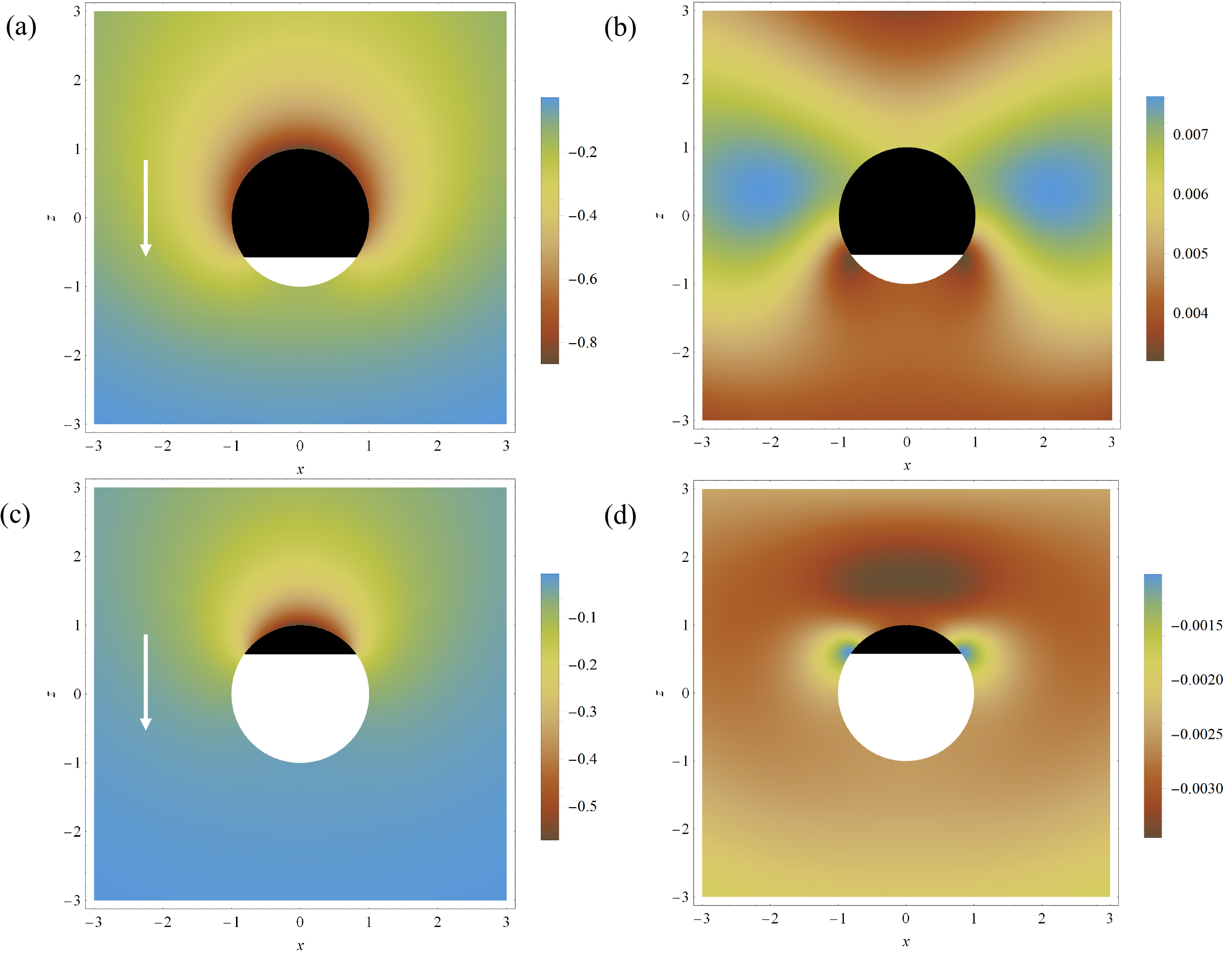}
        \caption{Relative solute concentration distribution around particle B and C in the Newtonian case (a) and (c) and second-order fluid contribution (b) and (d), respectively. The results are obtained for $Pe=2$ and $Da=0$ in the case of $M=-1$. The white arrow indicates the direction of motion.}
		\label{ConcMm1}
	\end{minipage}
\end{figure*}

To better understand this behaviour, the concentration profiles around the particles in a Newtonian fluid $c_0$ and the first-order contributions in a viscoelastic fluid $c_1$ are shown in Figures \ref{ConcA}, \ref{ConcMp1} and \ref{ConcMm1}. Note that in all cases the magnitude of the concentration perturbation field is very small, $\left|c_1\right|\ll \left|c_0\right|$, throughout the fluid; thus, even for order one Deborah numbers, the concentration field in a Newtonian fluid is virtually indistinguishable from that of a viscoelastic fluid (for $Pe=2$ and $Da=0$).

\begin{figure*}[t!]
\centering
		\begin{minipage}[t]{0.48\textwidth}
		        \includegraphics[width=\textwidth]{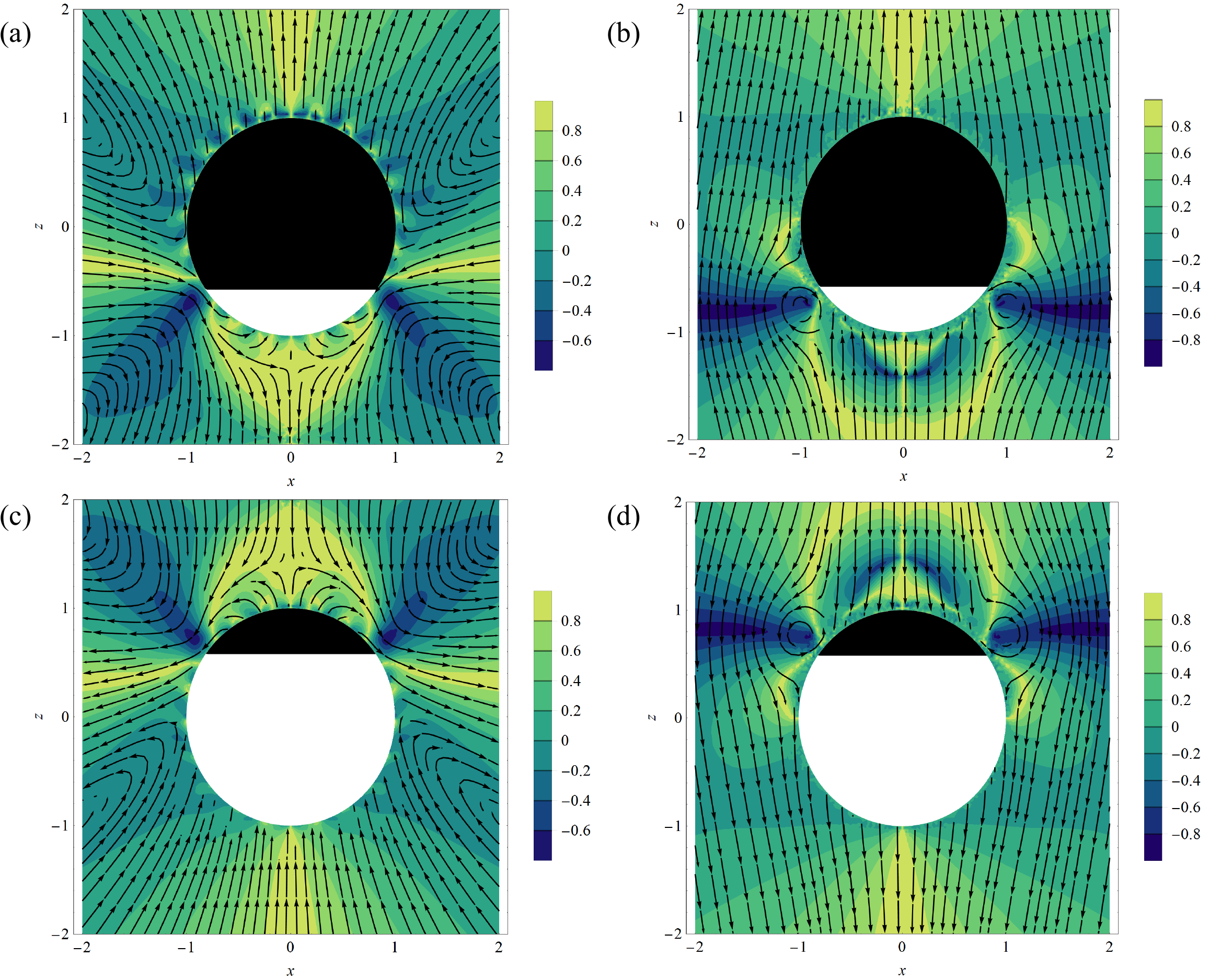}
        \caption{Stream plot of the flow field (lab frame) around particle B and C in the Newtonian case (a) and (c) and the second-order fluid first-order correction (b) and (d), respectively. The contour plots in the background represents the flow parameter for the different cases. The results are obtained for $Pe=2$ and $Da=0$ in the case of $M=1$.}
        \label{VelMp1}	
        \end{minipage}
    \hspace{10pt}
	\begin{minipage}[t]{0.48\textwidth}
        \includegraphics[width=\textwidth]{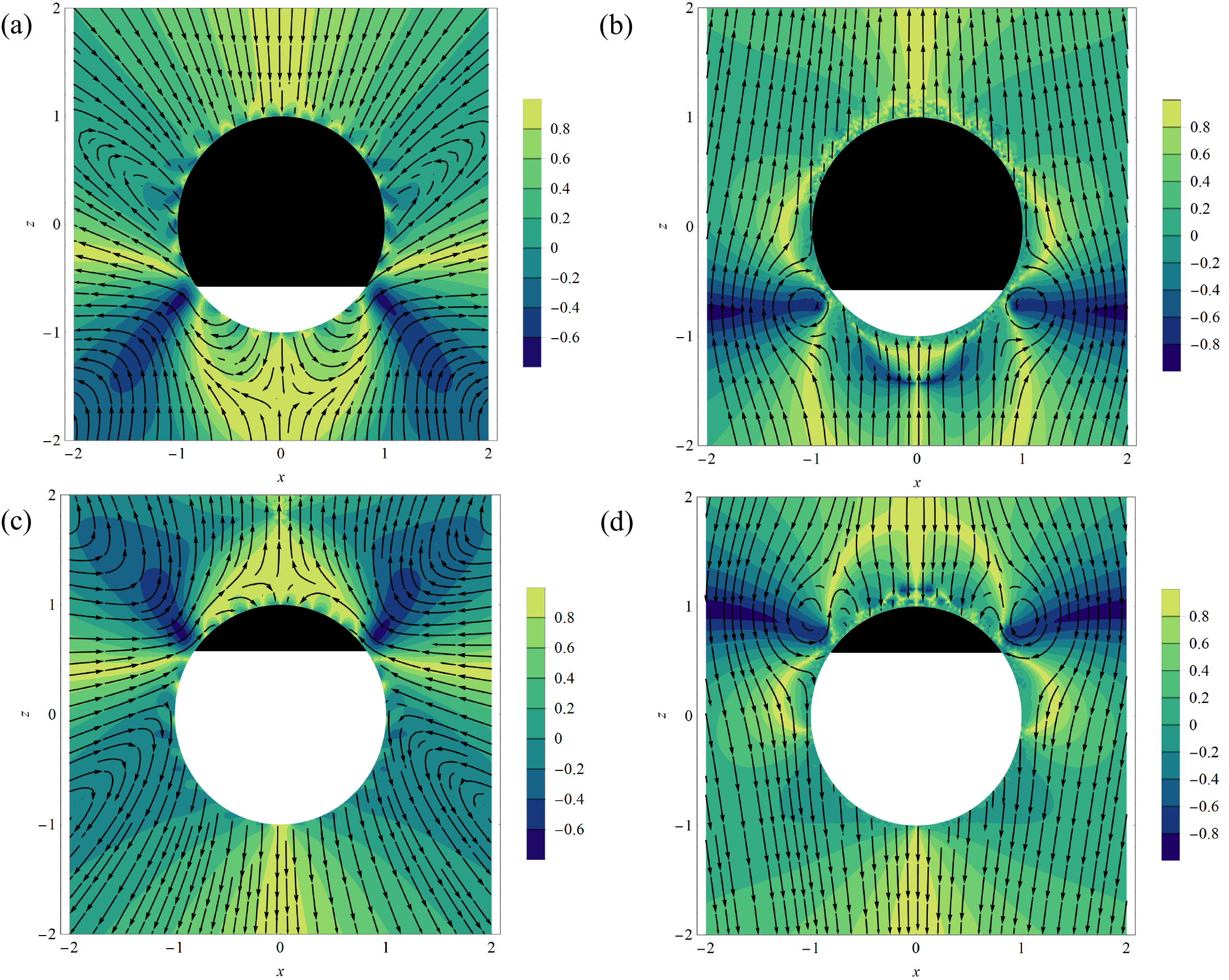}
        \caption{Stream plot of the flow field (lab frame) around particle B and C in the Newtonian case (a) and (c) and the second-order fluid first-order correction (b) and (d), respectively. The contour plots in the background represents the flow parameter for the different cases.  The results are obtained for $Pe=2$ and $Da=0$ in the case of $M=-1$.}
		\label{VelMm1}
	\end{minipage}
\end{figure*}

When the mobility is positive (negative), the particles swim towards their active (inert) pole. To leading order, this motion tends to spread (contract) the depleted region while also bringing a higher solute concentration towards the active (inert) cap (see Figure \ref{ConcA}a for $M=1$ and \ref{ConcA}c for $M=-1$). These effects both tend to reduce (enhance) the concentration difference between the front and back of the particle \citep{michelin14}. A reduction (enhancement) of the concentration gradient on the particle surface leads to a smaller (larger) surface slip velocity, and as a consequence, we see that propulsion velocities decrease (increase) with increasing of $Pe$ \citep{michelin14}. For particles with positive mobility, the velocities decrease monotonically. For particles with negative mobility, the velocity increase saturates, and at large P\'eclet a boundary layer develops while the speeds decay as $U \propto Pe^{-1/3}$ in a Newtonian fluid \citep{michelin14, yariv15}.

Inspecting the concentration perturbation field, a very different scenario appears (see Figure \ref{ConcA}b for $M=1$ and Figure \ref{ConcA}d for $M=-1$). The concentration perturbation (due to viscoelasticity) has a higher concentration in front of the swimmer, implying that the first-order slip velocity, given by the first term on the right-hand side of \eqref{Usof}, is opposite the leading-order term from the Newtonian result. However, given the magnitude of the concentration perturbation, the effect on the swimming speed is negligible. The second term on the right-hand side of \eqref{Usof}, gives the change in the velocity due to viscoelastic stresses. When $Pe=0$, this integral is identically zero for symmetric particles because the viscoelastic stresses induced on the particle are located symmetrically around the equator of the particles. Here, we find that this contribution is negligible for nonzero P\'eclet numbers as well, for particles with positive mobility. For particles with negative mobility, a moderate speed enhancement occurs at intermediate P\'eclet number due to viscoelastic stresses.

For asymmetric particles, the leading-order effect of the advection of the concentration field largely follows that of symmetric Janus particles (see Figure \ref{ConcMp1} for $M=1$ and Figure \ref{ConcMm1} for $M=-1$). Notably, for particles with negative mobility (see Figure \ref{ConcMm1}a and c), a larger active region leads to exacerbated gradient steepening and speedup while the opposite is true if the active region is smaller.

The effect of viscoelasticity on asymmetric particles is markedly different from the symmetric case as now the large viscoelastic stresses that arise at the discontinuity in surface activity are not symmetrically oriented at the equator of the particle. To illustrate, we show, in Figure \ref{VelMp1} ($M=1$) and Figure \ref{VelMm1} ($M=-1$), stream plots of the leading-order velocity field $\boldsymbol{u}_0$ and the velocity perturbation $\boldsymbol{u}_1$. We overlay these streamlines on contour plots of the flow type parameter $\chi$ \citep{patil2006constitutive} defined as follows 
\begin{align}\label{chi}
\chi=\frac{\lVert\dot{\boldsymbol{\gamma}}\rVert-\lVert\boldsymbol{\Theta}\rVert}{\lVert\dot{\boldsymbol{\gamma}}\rVert+\lVert\boldsymbol{\Theta}\rVert}
\end{align}
where $\boldsymbol{\Theta}=\boldsymbol{\nabla}\boldsymbol{u}-\left(\boldsymbol{\nabla}\boldsymbol{u}\right)^\top$ denotes the vorticity tensor. The flow type parameter $\chi$ ranges from -1 to 1 depending on the flow field. It assumes the value of 1 for pure extension, 0 for pure shear and -1 for pure rotation.

\begin{figure}[t!]
\centering
        \includegraphics[width=0.48\textwidth]{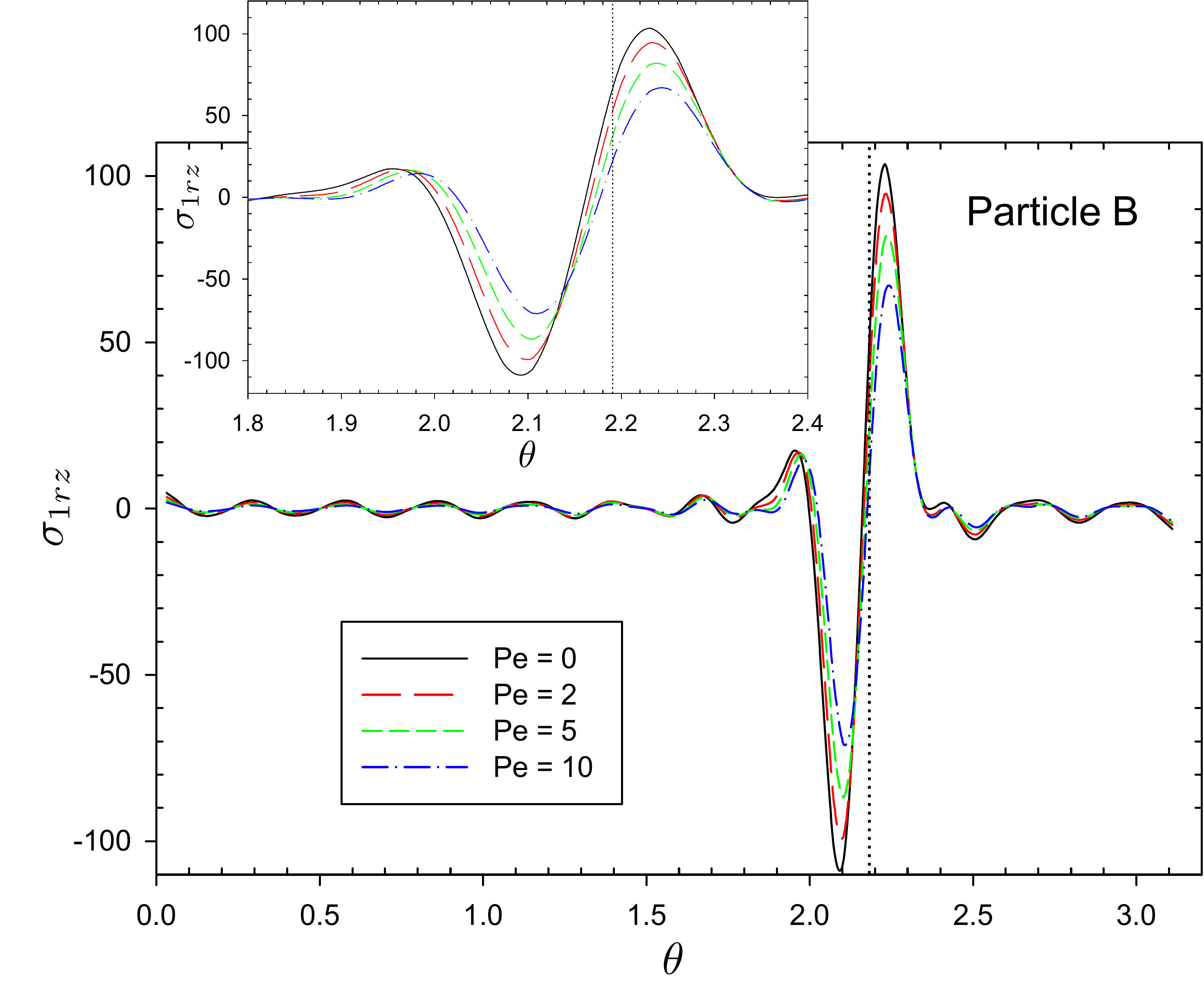}
        \caption{Viscoelastic stress on the surface of the particle of particle $B$ in the $\boldsymbol{e}_z$ direction, $\sigma_{1rz}$. The dashed vertical line indicates $\theta_c$. In this case, $M=1$, the magnitude of the viscoelastic stresses diminishes with increasing $Pe$ as gradients are smoothed by advection.}
		\label{stressfig}
\end{figure}

From the flow field plots, we see that the sharp gradient in the slip velocity at the discontinuity in surface activity gives rise to extensional flows on either side of the discontinuity and correspondingly large viscoelastic stresses that effectively push on the particle at this point, regardless of the sign of the mobility. In Figure \ref{stressfig} we show the viscoelastic stresses acting on the surface of the particle in the direction of motion, $\sigma_{1rz}$. Because these stresses are not located at the equator for asymmetric particles, this leads to a viscoelastic contribution to the speed given by the second term on the right-hand side of \eqref{Usof}. The effect of viscoelasticity can be easily predicted from this simple physical picture: if the activity discontinuity is on the back (front) of the particle with respect to the direction of motion the particle will see a speed increase (decrease) due to viscoelasticity. This description is also consistent with the literature where pusher-type squirmers see a speed increase in non-Newtonian fluids while puller-type squirmers see a speed decrease \citep{decorato15, datt2017active}, because for positive mobility particle B behaves as a pusher while particle C as a puller (and vice-versa for particles with negative mobility) \citep{datt2017active}. We note that \citet{zhu2012self} found numerically that the propulsion of pullers and pushers was always reduced in a viscoelastic fluid, modeled by the Giesekus model, due to the presence of large extensional stresses that develop to the aft of the swimmers at moderate Deborah numbers. These stresses are not significant at the small Deborah numbers captured by our approach but are certainly expected to grow as strains increase.

The effect of nonzero P\'eclet number on the viscoelastic stresses is more subtle, but can largely be explained from the physical picture given above. The effect of advection serves to reduce or enhance the concentration gradients at the surface of the particle which drive the slip boundary condition. The sharp gradients at the discontinuity of the surface activity also lead to the viscoelastic stresses that affect the speed of the particle. In Figure \ref{stressfig} we plot viscoelastic stresses acting on particle B with positive mobility. In this case, as the P\'eclet number increases, advection smooths the concentration gradients at the surface of the particle and indeed the viscoelastic stresses diminish. In general, if advection diminishes (enhances) the leading-order concentration gradients, as it does for positive (negative) mobility, then this in turn also diminishes (enhances) the viscoelastic stresses on the particle and leads to the changes in propulsion shown in Figure \ref{speedpos} (Figure \ref{speedneg}).

%%%%%%%%%%%%%%%%%%%%%%%%%%%%%%%%%%%%%%%%%%%%

\subsection{Reactive effect}
The effect of surface reactions follows straightforwardly from the physical picture developed in the preceding section. A reaction rate at the surface of the particle which depends linearly on the concentration will act to reduce the concentration gradients which drive surface slip. Because of this, increasing the Damk\"ohler number leads to a monotonic decrease in the propulsion speed of the particles \cite{michelin14}. Moreover, because those same concentration gradients drive the viscoelastic stresses on the particle, these stresses also decrease monotonically with increasing $Da$.

\section{Conclusions}
In this work, we investigated the dynamics of autophoretic Janus particles in weakly viscoelastic fluids. Using a combined asymptotic and numerical approach we solved for the concentration and flow fields for Janus particles with various surface boundary conditions immersed in a second-order fluid and investigated the impact of viscoelastity in the fluid to first order in Deborah number. The effect of viscoelasticity on the propulsion of the particle is largely due to the large viscoelastic stresses that are generated at the discontinuity in the surface coverage of the Janus particles. The discontinuity in surface activity drives a highly localized concentration variation near the surface of the particle and a sharp velocity gradient in the surface slip that results in large viscoelastic stresses. Advection can sharpen or diminish these concentration gradients and thus directly enhance or diminish the viscoelastic stresses, while concentration dependent chemical reactions act to diminish concentration gradients and thereby weaken viscoelastic effects on propulsion as well.

A limitation of our approach is that we consider only very small Deborah numbers due to our asymptotic approach to the viscoelasticity and use of the second-order fluid model, and so these conclusions may be significantly modified under large strains \citep{zhu2012self, decorato15, datt2017active}. On the flip side, this approach allows one to systematically disentangle the effects of advection and diffusion on viscoelastic stresses, and the picture may not be so clear at larger Deborah numbers. We also assume that the mobility is fixed; however, there is likely to be a first-order variation of the mobility coefficent as well for which one must undertake a matched asymptotic approach to determine \citep{michelin14}.

\bibliography{diffusiophoresis}

\begin{thebibliography}{56}%
\makeatletter
\providecommand \@ifxundefined [1]{%
 \@ifx{#1\undefined}
}%
\providecommand \@ifnum [1]{%
 \ifnum #1\expandafter \@firstoftwo
 \else \expandafter \@secondoftwo
 \fi
}%
\providecommand \@ifx [1]{%
 \ifx #1\expandafter \@firstoftwo
 \else \expandafter \@secondoftwo
 \fi
}%
\providecommand \natexlab [1]{#1}%
\providecommand \enquote  [1]{``#1''}%
\providecommand \bibnamefont  [1]{#1}%
\providecommand \bibfnamefont [1]{#1}%
\providecommand \citenamefont [1]{#1}%
\providecommand \href@noop [0]{\@secondoftwo}%
\providecommand \href [0]{\begingroup \@sanitize@url \@href}%
\providecommand \@href[1]{\@@startlink{#1}\@@href}%
\providecommand \@@href[1]{\endgroup#1\@@endlink}%
\providecommand \@sanitize@url [0]{\catcode `\\12\catcode `\$12\catcode
  `\&12\catcode `\#12\catcode `\^12\catcode `\_12\catcode `\%12\relax}%
\providecommand \@@startlink[1]{}%
\providecommand \@@endlink[0]{}%
\providecommand \url  [0]{\begingroup\@sanitize@url \@url }%
\providecommand \@url [1]{\endgroup\@href {#1}{\urlprefix }}%
\providecommand \urlprefix  [0]{URL }%
\providecommand \Eprint [0]{\href }%
\providecommand \doibase [0]{http://dx.doi.org/}%
\providecommand \selectlanguage [0]{\@gobble}%
\providecommand \bibinfo  [0]{\@secondoftwo}%
\providecommand \bibfield  [0]{\@secondoftwo}%
\providecommand \translation [1]{[#1]}%
\providecommand \BibitemOpen [0]{}%
\providecommand \bibitemStop [0]{}%
\providecommand \bibitemNoStop [0]{.\EOS\space}%
\providecommand \EOS [0]{\spacefactor3000\relax}%
\providecommand \BibitemShut  [1]{\csname bibitem#1\endcsname}%
\let\auto@bib@innerbib\@empty
%</preamble>
\bibitem [{\citenamefont {Ginot}\ \emph {et~al.}(2015)\citenamefont {Ginot},
  \citenamefont {Theurkauff}, \citenamefont {Levis}, \citenamefont {Ybert},
  \citenamefont {Bocquet}, \citenamefont {Berthier},\ and\ \citenamefont
  {Cottin-Bizonne}}]{Def_AC}%
  \BibitemOpen
  \bibfield  {author} {\bibinfo {author} {\bibfnamefont {F.}~\bibnamefont
  {Ginot}}, \bibinfo {author} {\bibfnamefont {I.}~\bibnamefont {Theurkauff}},
  \bibinfo {author} {\bibfnamefont {D.}~\bibnamefont {Levis}}, \bibinfo
  {author} {\bibfnamefont {C.}~\bibnamefont {Ybert}}, \bibinfo {author}
  {\bibfnamefont {L.}~\bibnamefont {Bocquet}}, \bibinfo {author} {\bibfnamefont
  {L.}~\bibnamefont {Berthier}}, \ and\ \bibinfo {author} {\bibfnamefont
  {C.}~\bibnamefont {Cottin-Bizonne}},\ }\href {\doibase
  10.1103/PhysRevX.5.011004} {\bibfield  {journal} {\bibinfo  {journal} {Phys.
  Rev. X}\ }\textbf {\bibinfo {volume} {5}},\ \bibinfo {pages} {011004}
  (\bibinfo {year} {2015})}\BibitemShut {NoStop}%
\bibitem [{\citenamefont {Nelson}\ \emph {et~al.}(2010)\citenamefont {Nelson},
  \citenamefont {Kaliakatsos},\ and\ \citenamefont
  {Abbott}}]{nelson2010microrobots}%
  \BibitemOpen
  \bibfield  {author} {\bibinfo {author} {\bibfnamefont {B.~J.}\ \bibnamefont
  {Nelson}}, \bibinfo {author} {\bibfnamefont {I.~K.}\ \bibnamefont
  {Kaliakatsos}}, \ and\ \bibinfo {author} {\bibfnamefont {J.~J.}\ \bibnamefont
  {Abbott}},\ }\href {\doibase 10.1146/annurev-bioeng-010510-103409} {\bibfield
   {journal} {\bibinfo  {journal} {Annu. Rev. Biomed. Eng.}\ }\textbf {\bibinfo
  {volume} {12}},\ \bibinfo {pages} {55} (\bibinfo {year} {2010})}\BibitemShut
  {NoStop}%
\bibitem [{\citenamefont {Soler}\ \emph {et~al.}(2013)\citenamefont {Soler},
  \citenamefont {Magdanz}, \citenamefont {Fomin}, \citenamefont {Sanchez},\
  and\ \citenamefont {Schmidt}}]{soler2013self}%
  \BibitemOpen
  \bibfield  {author} {\bibinfo {author} {\bibfnamefont {L.}~\bibnamefont
  {Soler}}, \bibinfo {author} {\bibfnamefont {V.}~\bibnamefont {Magdanz}},
  \bibinfo {author} {\bibfnamefont {V.~M.}\ \bibnamefont {Fomin}}, \bibinfo
  {author} {\bibfnamefont {S.}~\bibnamefont {Sanchez}}, \ and\ \bibinfo
  {author} {\bibfnamefont {O.~G.}\ \bibnamefont {Schmidt}},\ }\href {\doibase
  10.1021/nn405075d} {\bibfield  {journal} {\bibinfo  {journal} {{ACS} Nano}\
  }\textbf {\bibinfo {volume} {7}},\ \bibinfo {pages} {9611} (\bibinfo {year}
  {2013})}\BibitemShut {NoStop}%
\bibitem [{\citenamefont {Li}\ \emph {et~al.}(2014{\natexlab{a}})\citenamefont
  {Li}, \citenamefont {Singh}, \citenamefont {Sattayasamitsathit},
  \citenamefont {Orozco}, \citenamefont {Kaufmann}, \citenamefont {Dong},
  \citenamefont {Gao}, \citenamefont {Jurado-Sanchez}, \citenamefont
  {Fedorak},\ and\ \citenamefont {Wang}}]{li2014water}%
  \BibitemOpen
  \bibfield  {author} {\bibinfo {author} {\bibfnamefont {J.}~\bibnamefont
  {Li}}, \bibinfo {author} {\bibfnamefont {V.~V.}\ \bibnamefont {Singh}},
  \bibinfo {author} {\bibfnamefont {S.}~\bibnamefont {Sattayasamitsathit}},
  \bibinfo {author} {\bibfnamefont {J.}~\bibnamefont {Orozco}}, \bibinfo
  {author} {\bibfnamefont {K.}~\bibnamefont {Kaufmann}}, \bibinfo {author}
  {\bibfnamefont {R.}~\bibnamefont {Dong}}, \bibinfo {author} {\bibfnamefont
  {W.}~\bibnamefont {Gao}}, \bibinfo {author} {\bibfnamefont {B.}~\bibnamefont
  {Jurado-Sanchez}}, \bibinfo {author} {\bibfnamefont {Y.}~\bibnamefont
  {Fedorak}}, \ and\ \bibinfo {author} {\bibfnamefont {J.}~\bibnamefont
  {Wang}},\ }\href {\doibase 10.1021/nn505029k} {\bibfield  {journal} {\bibinfo
   {journal} {ACS Nano}\ }\textbf {\bibinfo {volume} {8}},\ \bibinfo {pages}
  {11118} (\bibinfo {year} {2014}{\natexlab{a}})}\BibitemShut {NoStop}%
\bibitem [{\citenamefont {Behrend}\ \emph {et~al.}(2004)\citenamefont
  {Behrend}, \citenamefont {Anker},\ and\ \citenamefont
  {Kopelman}}]{behrend2004brownian}%
  \BibitemOpen
  \bibfield  {author} {\bibinfo {author} {\bibfnamefont {C.~J.}\ \bibnamefont
  {Behrend}}, \bibinfo {author} {\bibfnamefont {J.~N.}\ \bibnamefont {Anker}},
  \ and\ \bibinfo {author} {\bibfnamefont {R.}~\bibnamefont {Kopelman}},\
  }\href {\doibase 10.1063/1.1637963} {\bibfield  {journal} {\bibinfo
  {journal} {Appl. Phys. Lett.}\ }\textbf {\bibinfo {volume} {84}},\ \bibinfo
  {pages} {154} (\bibinfo {year} {2004})}\BibitemShut {NoStop}%
\bibitem [{\citenamefont {Maggi}\ \emph {et~al.}(2015)\citenamefont {Maggi},
  \citenamefont {Simmchen}, \citenamefont {Saglimbeni}, \citenamefont {Katuri},
  \citenamefont {Dipalo}, \citenamefont {Angelis}, \citenamefont {Sanchez},\
  and\ \citenamefont {Leonardo}}]{maggi2016self}%
  \BibitemOpen
  \bibfield  {author} {\bibinfo {author} {\bibfnamefont {C.}~\bibnamefont
  {Maggi}}, \bibinfo {author} {\bibfnamefont {J.}~\bibnamefont {Simmchen}},
  \bibinfo {author} {\bibfnamefont {F.}~\bibnamefont {Saglimbeni}}, \bibinfo
  {author} {\bibfnamefont {J.}~\bibnamefont {Katuri}}, \bibinfo {author}
  {\bibfnamefont {M.}~\bibnamefont {Dipalo}}, \bibinfo {author} {\bibfnamefont
  {F.~D.}\ \bibnamefont {Angelis}}, \bibinfo {author} {\bibfnamefont
  {S.}~\bibnamefont {Sanchez}}, \ and\ \bibinfo {author} {\bibfnamefont
  {R.~D.}\ \bibnamefont {Leonardo}},\ }\href {\doibase 10.1002/smll.201502391}
  {\bibfield  {journal} {\bibinfo  {journal} {Small}\ }\textbf {\bibinfo
  {volume} {12}},\ \bibinfo {pages} {446} (\bibinfo {year} {2015})}\BibitemShut
  {NoStop}%
\bibitem [{\citenamefont {Z\"ottl}\ and\ \citenamefont
  {Stark}(2016)}]{Stark_2016}%
  \BibitemOpen
  \bibfield  {author} {\bibinfo {author} {\bibfnamefont {A.}~\bibnamefont
  {Z\"ottl}}\ and\ \bibinfo {author} {\bibfnamefont {H.}~\bibnamefont
  {Stark}},\ }\href {\doibase 10.1088/0953-8984/28/25/253001} {\bibfield
  {journal} {\bibinfo  {journal} {J. Phys.: Condens. Matter}\ }\textbf
  {\bibinfo {volume} {28}},\ \bibinfo {pages} {253001} (\bibinfo {year}
  {2016})}\BibitemShut {NoStop}%
\bibitem [{\citenamefont {Anderson}(1989)}]{anderson1989colloid}%
  \BibitemOpen
  \bibfield  {author} {\bibinfo {author} {\bibfnamefont {J.}~\bibnamefont
  {Anderson}},\ }\href {\doibase 10.1146/annurev.fluid.21.1.61} {\bibfield
  {journal} {\bibinfo  {journal} {Annu. Rev. Fluid Mech.}\ }\textbf {\bibinfo
  {volume} {21}},\ \bibinfo {pages} {61} (\bibinfo {year} {1989})}\BibitemShut
  {NoStop}%
\bibitem [{\citenamefont {Paxton}\ \emph {et~al.}(2004)\citenamefont {Paxton},
  \citenamefont {Kistler}, \citenamefont {Olmeda}, \citenamefont {Sen},
  \citenamefont {St.~Angelo}, \citenamefont {Cao}, \citenamefont {Mallouk},
  \citenamefont {Lammert},\ and\ \citenamefont {Crespi}}]{paxton2004catalytic}%
  \BibitemOpen
  \bibfield  {author} {\bibinfo {author} {\bibfnamefont {W.~F.}\ \bibnamefont
  {Paxton}}, \bibinfo {author} {\bibfnamefont {K.~C.}\ \bibnamefont {Kistler}},
  \bibinfo {author} {\bibfnamefont {C.~C.}\ \bibnamefont {Olmeda}}, \bibinfo
  {author} {\bibfnamefont {A.}~\bibnamefont {Sen}}, \bibinfo {author}
  {\bibfnamefont {S.~K.}\ \bibnamefont {St.~Angelo}}, \bibinfo {author}
  {\bibfnamefont {Y.}~\bibnamefont {Cao}}, \bibinfo {author} {\bibfnamefont
  {T.~E.}\ \bibnamefont {Mallouk}}, \bibinfo {author} {\bibfnamefont {P.~E.}\
  \bibnamefont {Lammert}}, \ and\ \bibinfo {author} {\bibfnamefont {V.~H.}\
  \bibnamefont {Crespi}},\ }\href {\doibase 10.1021/ja047697z} {\bibfield
  {journal} {\bibinfo  {journal} {J. Am. Chem. Soc.}\ }\textbf {\bibinfo
  {volume} {126}},\ \bibinfo {pages} {13424} (\bibinfo {year}
  {2004})}\BibitemShut {NoStop}%
\bibitem [{\citenamefont {C\'ordova-Figueroa}\ and\ \citenamefont
  {Brady}(2008)}]{cordova2008osmotic}%
  \BibitemOpen
  \bibfield  {author} {\bibinfo {author} {\bibfnamefont {U.~M.}\ \bibnamefont
  {C\'ordova-Figueroa}}\ and\ \bibinfo {author} {\bibfnamefont {J.~F.}\
  \bibnamefont {Brady}},\ }\href {\doibase 10.1103/PhysRevLett.100.158303}
  {\bibfield  {journal} {\bibinfo  {journal} {Phys. Rev. Lett.}\ }\textbf
  {\bibinfo {volume} {100}},\ \bibinfo {pages} {158303} (\bibinfo {year}
  {2008})}\BibitemShut {NoStop}%
\bibitem [{\citenamefont {Ebbens}\ and\ \citenamefont
  {Howse}(2011)}]{ebbens2011direct}%
  \BibitemOpen
  \bibfield  {author} {\bibinfo {author} {\bibfnamefont {S.~J.}\ \bibnamefont
  {Ebbens}}\ and\ \bibinfo {author} {\bibfnamefont {J.~R.}\ \bibnamefont
  {Howse}},\ }\href {\doibase 10.1021/la2033127} {\bibfield  {journal}
  {\bibinfo  {journal} {Langmuir}\ }\textbf {\bibinfo {volume} {27}},\ \bibinfo
  {pages} {12293} (\bibinfo {year} {2011})}\BibitemShut {NoStop}%
\bibitem [{\citenamefont {J\"ulicher}\ and\ \citenamefont
  {Prost}(2009)}]{julicher2009comment}%
  \BibitemOpen
  \bibfield  {author} {\bibinfo {author} {\bibfnamefont {F.}~\bibnamefont
  {J\"ulicher}}\ and\ \bibinfo {author} {\bibfnamefont {J.}~\bibnamefont
  {Prost}},\ }\href {\doibase 10.1103/PhysRevLett.103.079801} {\bibfield
  {journal} {\bibinfo  {journal} {Phys. Rev. Lett.}\ }\textbf {\bibinfo
  {volume} {103}},\ \bibinfo {pages} {079801} (\bibinfo {year}
  {2009})}\BibitemShut {NoStop}%
\bibitem [{\citenamefont {Golestanian}\ \emph {et~al.}(2005)\citenamefont
  {Golestanian}, \citenamefont {Liverpool},\ and\ \citenamefont
  {Ajdari}}]{golestanian2005propulsion}%
  \BibitemOpen
  \bibfield  {author} {\bibinfo {author} {\bibfnamefont {R.}~\bibnamefont
  {Golestanian}}, \bibinfo {author} {\bibfnamefont {T.~B.}\ \bibnamefont
  {Liverpool}}, \ and\ \bibinfo {author} {\bibfnamefont {A.}~\bibnamefont
  {Ajdari}},\ }\href {\doibase 10.1103/PhysRevLett.94.220801} {\bibfield
  {journal} {\bibinfo  {journal} {Phys. Rev. Lett.}\ }\textbf {\bibinfo
  {volume} {94}},\ \bibinfo {pages} {220801} (\bibinfo {year}
  {2005})}\BibitemShut {NoStop}%
\bibitem [{\citenamefont {Golestanian}\ \emph {et~al.}(2007)\citenamefont
  {Golestanian}, \citenamefont {Liverpool},\ and\ \citenamefont
  {Ajdari}}]{golestanian2007designing}%
  \BibitemOpen
  \bibfield  {author} {\bibinfo {author} {\bibfnamefont {R.}~\bibnamefont
  {Golestanian}}, \bibinfo {author} {\bibfnamefont {T.~B.}\ \bibnamefont
  {Liverpool}}, \ and\ \bibinfo {author} {\bibfnamefont {A.}~\bibnamefont
  {Ajdari}},\ }\href {http://stacks.iop.org/1367-2630/9/i=5/a=126} {\bibfield
  {journal} {\bibinfo  {journal} {New J. Phys.}\ }\textbf {\bibinfo {volume}
  {9}},\ \bibinfo {pages} {126} (\bibinfo {year} {2007})}\BibitemShut {NoStop}%
\bibitem [{\citenamefont {Walther}\ and\ \citenamefont
  {M{\"u}ller}(2013)}]{Janus_review}%
  \BibitemOpen
  \bibfield  {author} {\bibinfo {author} {\bibfnamefont {A.}~\bibnamefont
  {Walther}}\ and\ \bibinfo {author} {\bibfnamefont {A.~H.~E.}\ \bibnamefont
  {M{\"u}ller}},\ }\href {\doibase 10.1021/cr300089t} {\bibfield  {journal}
  {\bibinfo  {journal} {Chem. Rev.}\ }\textbf {\bibinfo {volume} {113}},\
  \bibinfo {pages} {5194} (\bibinfo {year} {2013})}\BibitemShut {NoStop}%
\bibitem [{\citenamefont {Mei}\ \emph {et~al.}(2011)\citenamefont {Mei},
  \citenamefont {Solovev}, \citenamefont {Sanchez},\ and\ \citenamefont
  {Schmidt}}]{mei2011rolled}%
  \BibitemOpen
  \bibfield  {author} {\bibinfo {author} {\bibfnamefont {Y.}~\bibnamefont
  {Mei}}, \bibinfo {author} {\bibfnamefont {A.~A.}\ \bibnamefont {Solovev}},
  \bibinfo {author} {\bibfnamefont {S.}~\bibnamefont {Sanchez}}, \ and\
  \bibinfo {author} {\bibfnamefont {O.~G.}\ \bibnamefont {Schmidt}},\ }\href
  {\doibase 10.1039/c0cs00078g} {\bibfield  {journal} {\bibinfo  {journal}
  {Chem. Soc. Rev.}\ }\textbf {\bibinfo {volume} {40}},\ \bibinfo {pages}
  {2109} (\bibinfo {year} {2011})}\BibitemShut {NoStop}%
\bibitem [{\citenamefont {Howse}\ \emph {et~al.}(2007)\citenamefont {Howse},
  \citenamefont {Jones}, \citenamefont {Ryan}, \citenamefont {Gough},
  \citenamefont {Vafabakhsh},\ and\ \citenamefont
  {Golestanian}}]{howse2007self}%
  \BibitemOpen
  \bibfield  {author} {\bibinfo {author} {\bibfnamefont {J.~R.}\ \bibnamefont
  {Howse}}, \bibinfo {author} {\bibfnamefont {R.~A.~L.}\ \bibnamefont {Jones}},
  \bibinfo {author} {\bibfnamefont {A.~J.}\ \bibnamefont {Ryan}}, \bibinfo
  {author} {\bibfnamefont {T.}~\bibnamefont {Gough}}, \bibinfo {author}
  {\bibfnamefont {R.}~\bibnamefont {Vafabakhsh}}, \ and\ \bibinfo {author}
  {\bibfnamefont {R.}~\bibnamefont {Golestanian}},\ }\href {\doibase
  10.1103/PhysRevLett.99.048102} {\bibfield  {journal} {\bibinfo  {journal}
  {Phys. Rev. Lett.}\ }\textbf {\bibinfo {volume} {99}},\ \bibinfo {pages}
  {048102} (\bibinfo {year} {2007})}\BibitemShut {NoStop}%
\bibitem [{\citenamefont {Ke}\ \emph {et~al.}(2010)\citenamefont {Ke},
  \citenamefont {Ye}, \citenamefont {Carroll},\ and\ \citenamefont
  {Showalter}}]{ke2010motion}%
  \BibitemOpen
  \bibfield  {author} {\bibinfo {author} {\bibfnamefont {H.}~\bibnamefont
  {Ke}}, \bibinfo {author} {\bibfnamefont {S.}~\bibnamefont {Ye}}, \bibinfo
  {author} {\bibfnamefont {R.~L.}\ \bibnamefont {Carroll}}, \ and\ \bibinfo
  {author} {\bibfnamefont {K.}~\bibnamefont {Showalter}},\ }\href {\doibase
  10.1021/jp101193u} {\bibfield  {journal} {\bibinfo  {journal} {J. Phys. Chem.
  A}\ }\textbf {\bibinfo {volume} {114}},\ \bibinfo {pages} {5462} (\bibinfo
  {year} {2010})}\BibitemShut {NoStop}%
\bibitem [{\citenamefont {Campbell}\ and\ \citenamefont
  {Ebbens}(2013)}]{campbell2013gravitaxis}%
  \BibitemOpen
  \bibfield  {author} {\bibinfo {author} {\bibfnamefont {A.~I.}\ \bibnamefont
  {Campbell}}\ and\ \bibinfo {author} {\bibfnamefont {S.~J.}\ \bibnamefont
  {Ebbens}},\ }\href {\doibase 10.1021/la403450j} {\bibfield  {journal}
  {\bibinfo  {journal} {Langmuir}\ }\textbf {\bibinfo {volume} {29}},\ \bibinfo
  {pages} {14066} (\bibinfo {year} {2013})}\BibitemShut {NoStop}%
\bibitem [{\citenamefont {Brown}\ and\ \citenamefont
  {Poon}(2014)}]{brown2014ionic}%
  \BibitemOpen
  \bibfield  {author} {\bibinfo {author} {\bibfnamefont {A.}~\bibnamefont
  {Brown}}\ and\ \bibinfo {author} {\bibfnamefont {W.}~\bibnamefont {Poon}},\
  }\href {\doibase 10.1039/c4sm00340c} {\bibfield  {journal} {\bibinfo
  {journal} {Soft Matt.}\ }\textbf {\bibinfo {volume} {10}},\ \bibinfo {pages}
  {4016} (\bibinfo {year} {2014})}\BibitemShut {NoStop}%
\bibitem [{\citenamefont {Brown}\ \emph {et~al.}(2017)\citenamefont {Brown},
  \citenamefont {Poon}, \citenamefont {Holm},\ and\ \citenamefont
  {de~Graaf}}]{brown2015bulk}%
  \BibitemOpen
  \bibfield  {author} {\bibinfo {author} {\bibfnamefont {A.~T.}\ \bibnamefont
  {Brown}}, \bibinfo {author} {\bibfnamefont {W.~C.~K.}\ \bibnamefont {Poon}},
  \bibinfo {author} {\bibfnamefont {C.}~\bibnamefont {Holm}}, \ and\ \bibinfo
  {author} {\bibfnamefont {J.}~\bibnamefont {de~Graaf}},\ }\href {\doibase
  10.1039/c6sm01867j} {\bibfield  {journal} {\bibinfo  {journal} {Soft Matt.}\
  }\textbf {\bibinfo {volume} {13}},\ \bibinfo {pages} {1200} (\bibinfo {year}
  {2017})}\BibitemShut {NoStop}%
\bibitem [{\citenamefont {Jiang}\ \emph {et~al.}(2010)\citenamefont {Jiang},
  \citenamefont {Yoshinaga},\ and\ \citenamefont {Sano}}]{jiang2010active}%
  \BibitemOpen
  \bibfield  {author} {\bibinfo {author} {\bibfnamefont {H.-R.}\ \bibnamefont
  {Jiang}}, \bibinfo {author} {\bibfnamefont {N.}~\bibnamefont {Yoshinaga}}, \
  and\ \bibinfo {author} {\bibfnamefont {M.}~\bibnamefont {Sano}},\ }\href
  {\doibase 10.1103/PhysRevLett.105.268302} {\bibfield  {journal} {\bibinfo
  {journal} {Phys. Rev. Lett.}\ }\textbf {\bibinfo {volume} {105}},\ \bibinfo
  {pages} {268302} (\bibinfo {year} {2010})}\BibitemShut {NoStop}%
\bibitem [{\citenamefont {Moran}\ and\ \citenamefont
  {Posner}(2017)}]{Moran_2017}%
  \BibitemOpen
  \bibfield  {author} {\bibinfo {author} {\bibfnamefont {J.~L.}\ \bibnamefont
  {Moran}}\ and\ \bibinfo {author} {\bibfnamefont {J.~D.}\ \bibnamefont
  {Posner}},\ }\href {\doibase 10.1146/annurev-fluid-122414-034456} {\bibfield
  {journal} {\bibinfo  {journal} {Annu. Rev. Fluid Mech.}\ }\textbf {\bibinfo
  {volume} {49}},\ \bibinfo {pages} {511} (\bibinfo {year} {2017})}\BibitemShut
  {NoStop}%
\bibitem [{\citenamefont {J{\"u}licher}\ and\ \citenamefont
  {Prost}(2009)}]{julicher2009generic}%
  \BibitemOpen
  \bibfield  {author} {\bibinfo {author} {\bibfnamefont {F.}~\bibnamefont
  {J{\"u}licher}}\ and\ \bibinfo {author} {\bibfnamefont {J.}~\bibnamefont
  {Prost}},\ }\href {\doibase 10.1140/epje/i2008-10446-8} {\bibfield  {journal}
  {\bibinfo  {journal} {Eur. Phys. J. E}\ }\textbf {\bibinfo {volume} {29}},\
  \bibinfo {pages} {27} (\bibinfo {year} {2009})}\BibitemShut {NoStop}%
\bibitem [{\citenamefont {C{\'{o}}rdova-Figueroa}\ \emph
  {et~al.}(2013)\citenamefont {C{\'{o}}rdova-Figueroa}, \citenamefont {Brady},\
  and\ \citenamefont {Shklyaev}}]{cordova2013osmotic}%
  \BibitemOpen
  \bibfield  {author} {\bibinfo {author} {\bibfnamefont {U.~M.}\ \bibnamefont
  {C{\'{o}}rdova-Figueroa}}, \bibinfo {author} {\bibfnamefont {J.~F.}\
  \bibnamefont {Brady}}, \ and\ \bibinfo {author} {\bibfnamefont
  {S.}~\bibnamefont {Shklyaev}},\ }\href {\doibase 10.1039/c3sm00017f}
  {\bibfield  {journal} {\bibinfo  {journal} {Soft Matt.}\ }\textbf {\bibinfo
  {volume} {9}},\ \bibinfo {pages} {6382} (\bibinfo {year} {2013})}\BibitemShut
  {NoStop}%
\bibitem [{\citenamefont {Popescu}\ \emph {et~al.}(2010)\citenamefont
  {Popescu}, \citenamefont {Dietrich}, \citenamefont {Tasinkevych},\ and\
  \citenamefont {Ralston}}]{popescu2010phoretic}%
  \BibitemOpen
  \bibfield  {author} {\bibinfo {author} {\bibfnamefont {M.~N.}\ \bibnamefont
  {Popescu}}, \bibinfo {author} {\bibfnamefont {S.}~\bibnamefont {Dietrich}},
  \bibinfo {author} {\bibfnamefont {M.}~\bibnamefont {Tasinkevych}}, \ and\
  \bibinfo {author} {\bibfnamefont {J.}~\bibnamefont {Ralston}},\ }\href
  {\doibase 10.1140/epje/i2010-10593-3} {\bibfield  {journal} {\bibinfo
  {journal} {Eur. Phys. J. E}\ }\textbf {\bibinfo {volume} {31}},\ \bibinfo
  {pages} {351} (\bibinfo {year} {2010})}\BibitemShut {NoStop}%
\bibitem [{\citenamefont {Ebbens}\ \emph {et~al.}(2012)\citenamefont {Ebbens},
  \citenamefont {Tu}, \citenamefont {Howse},\ and\ \citenamefont
  {Golestanian}}]{ebbens2012size}%
  \BibitemOpen
  \bibfield  {author} {\bibinfo {author} {\bibfnamefont {S.}~\bibnamefont
  {Ebbens}}, \bibinfo {author} {\bibfnamefont {M.-H.}\ \bibnamefont {Tu}},
  \bibinfo {author} {\bibfnamefont {J.~R.}\ \bibnamefont {Howse}}, \ and\
  \bibinfo {author} {\bibfnamefont {R.}~\bibnamefont {Golestanian}},\ }\href
  {\doibase 10.1103/PhysRevE.85.020401} {\bibfield  {journal} {\bibinfo
  {journal} {Phys. Rev. E}\ }\textbf {\bibinfo {volume} {85}},\ \bibinfo
  {pages} {020401} (\bibinfo {year} {2012})}\BibitemShut {NoStop}%
\bibitem [{\citenamefont {Sabass}\ and\ \citenamefont
  {Seifert}(2012)}]{sabass2012dynamics}%
  \BibitemOpen
  \bibfield  {author} {\bibinfo {author} {\bibfnamefont {B.}~\bibnamefont
  {Sabass}}\ and\ \bibinfo {author} {\bibfnamefont {U.}~\bibnamefont
  {Seifert}},\ }\href {\doibase 10.1063/1.3681143} {\bibfield  {journal}
  {\bibinfo  {journal} {J. Chem. Phys.}\ }\textbf {\bibinfo {volume} {136}},\
  \bibinfo {pages} {064508} (\bibinfo {year} {2012})}\BibitemShut {NoStop}%
\bibitem [{\citenamefont {Sharifi-Mood}\ \emph {et~al.}(2013)\citenamefont
  {Sharifi-Mood}, \citenamefont {Koplik},\ and\ \citenamefont
  {Maldarelli}}]{sharifi2013diffusiophoretic}%
  \BibitemOpen
  \bibfield  {author} {\bibinfo {author} {\bibfnamefont {N.}~\bibnamefont
  {Sharifi-Mood}}, \bibinfo {author} {\bibfnamefont {J.}~\bibnamefont
  {Koplik}}, \ and\ \bibinfo {author} {\bibfnamefont {C.}~\bibnamefont
  {Maldarelli}},\ }\href {\doibase 10.1063/1.4772978} {\bibfield  {journal}
  {\bibinfo  {journal} {Phys. Fluids}\ }\textbf {\bibinfo {volume} {25}},\
  \bibinfo {pages} {012001} (\bibinfo {year} {2013})}\BibitemShut {NoStop}%
\bibitem [{\citenamefont {Michelin}\ and\ \citenamefont
  {Lauga}(2014)}]{michelin14}%
  \BibitemOpen
  \bibfield  {author} {\bibinfo {author} {\bibfnamefont {S.}~\bibnamefont
  {Michelin}}\ and\ \bibinfo {author} {\bibfnamefont {E.}~\bibnamefont
  {Lauga}},\ }\href {\doibase 10.1017/jfm.2014.158} {\bibfield  {journal}
  {\bibinfo  {journal} {J. Fluid Mech.}\ }\textbf {\bibinfo {volume} {747}},\
  \bibinfo {pages} {572} (\bibinfo {year} {2014})}\BibitemShut {NoStop}%
\bibitem [{\citenamefont {Patteson}\ \emph {et~al.}(2016)\citenamefont
  {Patteson}, \citenamefont {Gopinath},\ and\ \citenamefont
  {Arratia}}]{arratia_review}%
  \BibitemOpen
  \bibfield  {author} {\bibinfo {author} {\bibfnamefont {A.~E.}\ \bibnamefont
  {Patteson}}, \bibinfo {author} {\bibfnamefont {A.}~\bibnamefont {Gopinath}},
  \ and\ \bibinfo {author} {\bibfnamefont {P.~E.}\ \bibnamefont {Arratia}},\
  }\href {\doibase http://dx.doi.org/10.1016/j.cocis.2016.01.001} {\bibfield
  {journal} {\bibinfo  {journal} {Curr. Opin. Colloid Interface Sci.}\ }\textbf
  {\bibinfo {volume} {21}},\ \bibinfo {pages} {86} (\bibinfo {year}
  {2016})}\BibitemShut {NoStop}%
\bibitem [{\citenamefont {Datt}\ \emph {et~al.}(2017)\citenamefont {Datt},
  \citenamefont {Natale}, \citenamefont {Hatzikiriakos},\ and\ \citenamefont
  {Elfring}}]{datt2017active}%
  \BibitemOpen
  \bibfield  {author} {\bibinfo {author} {\bibfnamefont {C.}~\bibnamefont
  {Datt}}, \bibinfo {author} {\bibfnamefont {G.}~\bibnamefont {Natale}},
  \bibinfo {author} {\bibfnamefont {S.~G.}\ \bibnamefont {Hatzikiriakos}}, \
  and\ \bibinfo {author} {\bibfnamefont {G.~J.}\ \bibnamefont {Elfring}},\
  }\href {\doibase 10.1017/jfm.2017.353} {\bibfield  {journal} {\bibinfo
  {journal} {J. Fluid Mech.}\ }\textbf {\bibinfo {volume} {823}},\ \bibinfo
  {pages} {675–688} (\bibinfo {year} {2017})}\BibitemShut {NoStop}%
\bibitem [{\citenamefont {Oppenheimer}\ \emph {et~al.}(2016)\citenamefont
  {Oppenheimer}, \citenamefont {Navardi},\ and\ \citenamefont
  {Stone}}]{PRFStone}%
  \BibitemOpen
  \bibfield  {author} {\bibinfo {author} {\bibfnamefont {N.}~\bibnamefont
  {Oppenheimer}}, \bibinfo {author} {\bibfnamefont {S.}~\bibnamefont
  {Navardi}}, \ and\ \bibinfo {author} {\bibfnamefont {H.~A.}\ \bibnamefont
  {Stone}},\ }\href {\doibase 10.1103/PhysRevFluids.1.014001} {\bibfield
  {journal} {\bibinfo  {journal} {Phys. Rev. Fluids}\ }\textbf {\bibinfo
  {volume} {1}},\ \bibinfo {pages} {014001} (\bibinfo {year}
  {2016})}\BibitemShut {NoStop}%
\bibitem [{\citenamefont {Gomez-Solano}\ \emph {et~al.}(2016)\citenamefont
  {Gomez-Solano}, \citenamefont {Blokhuis},\ and\ \citenamefont
  {Bechinger}}]{gomez2016dynamics}%
  \BibitemOpen
  \bibfield  {author} {\bibinfo {author} {\bibfnamefont {J.~R.}\ \bibnamefont
  {Gomez-Solano}}, \bibinfo {author} {\bibfnamefont {A.}~\bibnamefont
  {Blokhuis}}, \ and\ \bibinfo {author} {\bibfnamefont {C.}~\bibnamefont
  {Bechinger}},\ }\href {\doibase 10.1103/PhysRevLett.116.138301} {\bibfield
  {journal} {\bibinfo  {journal} {Phys. Rev. Lett.}\ }\textbf {\bibinfo
  {volume} {116}},\ \bibinfo {pages} {138301} (\bibinfo {year}
  {2016})}\BibitemShut {NoStop}%
\bibitem [{\citenamefont {Elfring}\ and\ \citenamefont
  {Lauga}(2015)}]{elfring14}%
  \BibitemOpen
  \bibfield  {author} {\bibinfo {author} {\bibfnamefont {G.~J.}\ \bibnamefont
  {Elfring}}\ and\ \bibinfo {author} {\bibfnamefont {E.}~\bibnamefont
  {Lauga}},\ }\enquote {\bibinfo {title} {Theory of locomotion through complex
  fluids},}\ in\ \href {\doibase 10.1007/978-1-4939-2065-5_8} {\emph {\bibinfo
  {booktitle} {Complex Fluids in Biological Systems: Experiment, Theory, and
  Computation}}},\ \bibinfo {editor} {edited by\ \bibinfo {editor}
  {\bibfnamefont {S.~E.}\ \bibnamefont {Spagnolie}}}\ (\bibinfo  {publisher}
  {Springer New York},\ \bibinfo {address} {New York, NY},\ \bibinfo {year}
  {2015})\ pp.\ \bibinfo {pages} {283--317}\BibitemShut {NoStop}%
\bibitem [{\citenamefont {Sznitman}\ and\ \citenamefont
  {Arratia}(2015)}]{sznitman2015}%
  \BibitemOpen
  \bibfield  {author} {\bibinfo {author} {\bibfnamefont {J.}~\bibnamefont
  {Sznitman}}\ and\ \bibinfo {author} {\bibfnamefont {P.~E.}\ \bibnamefont
  {Arratia}},\ }in\ \href {\doibase 10.1007/978-1-4939-2065-5_7} {\emph
  {\bibinfo {booktitle} {Complex Fluids in Biological Systems: Experiment,
  Theory, and Computation}}},\ \bibinfo {editor} {edited by\ \bibinfo {editor}
  {\bibfnamefont {E.~S.}\ \bibnamefont {Spagnolie}}}\ (\bibinfo  {publisher}
  {Springer New York},\ \bibinfo {address} {New York, NY},\ \bibinfo {year}
  {2015})\ Chap.\ \bibinfo {chapter} {Locomotion Through Complex Fluids: An
  Experimental View}, pp.\ \bibinfo {pages} {245--281}\BibitemShut {NoStop}%
\bibitem [{\citenamefont {Zhu}\ \emph {et~al.}(2012)\citenamefont {Zhu},
  \citenamefont {Lauga},\ and\ \citenamefont {Brandt}}]{zhu2012self}%
  \BibitemOpen
  \bibfield  {author} {\bibinfo {author} {\bibfnamefont {L.}~\bibnamefont
  {Zhu}}, \bibinfo {author} {\bibfnamefont {E.}~\bibnamefont {Lauga}}, \ and\
  \bibinfo {author} {\bibfnamefont {L.}~\bibnamefont {Brandt}},\ }\href
  {\doibase http://dx.doi.org/10.1063/1.4718446} {\bibfield  {journal}
  {\bibinfo  {journal} {Phys. Fluids}\ }\textbf {\bibinfo {volume} {24}},\
  \bibinfo {pages} {051902} (\bibinfo {year} {2012})}\BibitemShut {NoStop}%
\bibitem [{\citenamefont {Montenegro-Johnson}\ \emph
  {et~al.}(2013)\citenamefont {Montenegro-Johnson}, \citenamefont {Smith},\
  and\ \citenamefont {Loghin}}]{montenegro}%
  \BibitemOpen
  \bibfield  {author} {\bibinfo {author} {\bibfnamefont {T.~D.}\ \bibnamefont
  {Montenegro-Johnson}}, \bibinfo {author} {\bibfnamefont {D.~J.}\ \bibnamefont
  {Smith}}, \ and\ \bibinfo {author} {\bibfnamefont {D.}~\bibnamefont
  {Loghin}},\ }\href {\doibase http://dx.doi.org/10.1063/1.4818640} {\bibfield
  {journal} {\bibinfo  {journal} {Phys. Fluids}\ }\textbf {\bibinfo {volume}
  {25}},\ \bibinfo {pages} {081903} (\bibinfo {year} {2013})}\BibitemShut
  {NoStop}%
\bibitem [{\citenamefont {Li}\ \emph {et~al.}(2014{\natexlab{b}})\citenamefont
  {Li}, \citenamefont {Karimi},\ and\ \citenamefont {Ardekani}}]{Li2014}%
  \BibitemOpen
  \bibfield  {author} {\bibinfo {author} {\bibfnamefont {G.~J.}\ \bibnamefont
  {Li}}, \bibinfo {author} {\bibfnamefont {A.}~\bibnamefont {Karimi}}, \ and\
  \bibinfo {author} {\bibfnamefont {A.~M.}\ \bibnamefont {Ardekani}},\ }\href
  {\doibase 10.1007/s00397-014-0796-9} {\bibfield  {journal} {\bibinfo
  {journal} {Rheol. Acta}\ }\textbf {\bibinfo {volume} {53}},\ \bibinfo {pages}
  {911} (\bibinfo {year} {2014}{\natexlab{b}})}\BibitemShut {NoStop}%
\bibitem [{\citenamefont {De~Corato}\ \emph {et~al.}(2015)\citenamefont
  {De~Corato}, \citenamefont {Greco},\ and\ \citenamefont
  {Maffettone}}]{decorato15}%
  \BibitemOpen
  \bibfield  {author} {\bibinfo {author} {\bibfnamefont {M.}~\bibnamefont
  {De~Corato}}, \bibinfo {author} {\bibfnamefont {F.}~\bibnamefont {Greco}}, \
  and\ \bibinfo {author} {\bibfnamefont {P.~L.}\ \bibnamefont {Maffettone}},\
  }\href {\doibase 10.1103/PhysRevE.92.053008} {\bibfield  {journal} {\bibinfo
  {journal} {Phys. Rev. E}\ }\textbf {\bibinfo {volume} {92}},\ \bibinfo
  {pages} {053008} (\bibinfo {year} {2015})}\BibitemShut {NoStop}%
\bibitem [{\citenamefont {Datt}\ \emph {et~al.}(2015)\citenamefont {Datt},
  \citenamefont {Zhu}, \citenamefont {Elfring},\ and\ \citenamefont
  {Pak}}]{jfmCharu}%
  \BibitemOpen
  \bibfield  {author} {\bibinfo {author} {\bibfnamefont {C.}~\bibnamefont
  {Datt}}, \bibinfo {author} {\bibfnamefont {L.}~\bibnamefont {Zhu}}, \bibinfo
  {author} {\bibfnamefont {G.~J.}\ \bibnamefont {Elfring}}, \ and\ \bibinfo
  {author} {\bibfnamefont {O.~S.}\ \bibnamefont {Pak}},\ }\href {\doibase
  10.1017/jfm.2015.600} {\bibfield  {journal} {\bibinfo  {journal} {J. Fluid
  Mech.}\ }\textbf {\bibinfo {volume} {784}},\ \bibinfo {pages} {R1} (\bibinfo
  {year} {2015})}\BibitemShut {NoStop}%
\bibitem [{\citenamefont {Lighthill}(1952)}]{Lighthill1951}%
  \BibitemOpen
  \bibfield  {author} {\bibinfo {author} {\bibfnamefont {M.~J.}\ \bibnamefont
  {Lighthill}},\ }\href {\doibase 10.1002/cpa.3160050201} {\bibfield  {journal}
  {\bibinfo  {journal} {Commun. Pure Appl. Math.}\ }\textbf {\bibinfo {volume}
  {5}},\ \bibinfo {pages} {109} (\bibinfo {year} {1952})}\BibitemShut {NoStop}%
\bibitem [{\citenamefont {Pedley}(2016)}]{Pedley_review}%
  \BibitemOpen
  \bibfield  {author} {\bibinfo {author} {\bibfnamefont {T.~J.}\ \bibnamefont
  {Pedley}},\ }\href {\doibase 10.1093/imamat/hxw030} {\bibfield  {journal}
  {\bibinfo  {journal} {IMA J. Appl. Math.}\ }\textbf {\bibinfo {volume}
  {81}},\ \bibinfo {pages} {488} (\bibinfo {year} {2016})}\BibitemShut
  {NoStop}%
\bibitem [{\citenamefont {Bird}\ \emph {et~al.}(1977)\citenamefont {Bird},
  \citenamefont {Armstrong}, \citenamefont {Hassager},\ and\ \citenamefont
  {Curtiss}}]{bird1977dynamics}%
  \BibitemOpen
  \bibfield  {author} {\bibinfo {author} {\bibfnamefont {R.~B.}\ \bibnamefont
  {Bird}}, \bibinfo {author} {\bibfnamefont {R.~C.}\ \bibnamefont {Armstrong}},
  \bibinfo {author} {\bibfnamefont {O.}~\bibnamefont {Hassager}}, \ and\
  \bibinfo {author} {\bibfnamefont {C.~F.}\ \bibnamefont {Curtiss}},\
  }\href@noop {} {\emph {\bibinfo {title} {Dynamics of polymeric liquids}}},\
  Vol.~\bibinfo {volume} {1}\ (\bibinfo  {publisher} {Wiley New York},\
  \bibinfo {year} {1977})\BibitemShut {NoStop}%
\bibitem [{\citenamefont {De~Corato}\ \emph {et~al.}(2016)\citenamefont
  {De~Corato}, \citenamefont {Greco},\ and\ \citenamefont
  {Maffettone}}]{decorato16b}%
  \BibitemOpen
  \bibfield  {author} {\bibinfo {author} {\bibfnamefont {M.}~\bibnamefont
  {De~Corato}}, \bibinfo {author} {\bibfnamefont {F.}~\bibnamefont {Greco}}, \
  and\ \bibinfo {author} {\bibfnamefont {P.~L.}\ \bibnamefont {Maffettone}},\
  }\href {\doibase 10.1103/PhysRevE.94.057102} {\bibfield  {journal} {\bibinfo
  {journal} {Phys. Rev. E}\ }\textbf {\bibinfo {volume} {94}},\ \bibinfo
  {pages} {057102} (\bibinfo {year} {2016})}\BibitemShut {NoStop}%
\bibitem [{\citenamefont {Bird}\ and\ \citenamefont {Wiest}(1995)}]{bird95}%
  \BibitemOpen
  \bibfield  {author} {\bibinfo {author} {\bibfnamefont {R.~B.}\ \bibnamefont
  {Bird}}\ and\ \bibinfo {author} {\bibfnamefont {J.~M.}\ \bibnamefont
  {Wiest}},\ }\href {\doibase 10.1146/annurev.fl.27.010195.001125} {\bibfield
  {journal} {\bibinfo  {journal} {Ann. Rev. Fluid Mech.}\ }\textbf {\bibinfo
  {volume} {27}},\ \bibinfo {pages} {169} (\bibinfo {year} {1995})}\BibitemShut
  {NoStop}%
\bibitem [{\citenamefont {Leal}(1980)}]{leal80}%
  \BibitemOpen
  \bibfield  {author} {\bibinfo {author} {\bibfnamefont {L.~G.}\ \bibnamefont
  {Leal}},\ }\href {\doibase 10.1146/annurev.fl.12.010180.002251} {\bibfield
  {journal} {\bibinfo  {journal} {Ann. Rev. Fluid Mech.}\ }\textbf {\bibinfo
  {volume} {12}},\ \bibinfo {pages} {435} (\bibinfo {year} {1980})}\BibitemShut
  {NoStop}%
\bibitem [{\citenamefont {Happel}\ and\ \citenamefont
  {Brenner}(1965)}]{happel65}%
  \BibitemOpen
  \bibfield  {author} {\bibinfo {author} {\bibfnamefont {J.}~\bibnamefont
  {Happel}}\ and\ \bibinfo {author} {\bibfnamefont {H.}~\bibnamefont
  {Brenner}},\ }\href@noop {} {\emph {\bibinfo {title} {Low {R}eynolds {N}umber
  {H}ydrodynamics}}}\ (\bibinfo  {publisher} {Prentice-Hall, Inc.},\ \bibinfo
  {year} {1965})\BibitemShut {NoStop}%
\bibitem [{\citenamefont {Stone}\ and\ \citenamefont {Samuel}(1996)}]{stone96}%
  \BibitemOpen
  \bibfield  {author} {\bibinfo {author} {\bibfnamefont {H.~A.}\ \bibnamefont
  {Stone}}\ and\ \bibinfo {author} {\bibfnamefont {A.~D.~T.}\ \bibnamefont
  {Samuel}},\ }\href {\doibase 10.1103/PhysRevLett.77.4102} {\bibfield
  {journal} {\bibinfo  {journal} {Phys. Rev. Lett.}\ }\textbf {\bibinfo
  {volume} {77}},\ \bibinfo {pages} {4102} (\bibinfo {year}
  {1996})}\BibitemShut {NoStop}%
\bibitem [{\citenamefont {Lauga}(2009)}]{lauga09}%
  \BibitemOpen
  \bibfield  {author} {\bibinfo {author} {\bibfnamefont {E.}~\bibnamefont
  {Lauga}},\ }\href {http://stacks.iop.org/0295-5075/86/i=6/a=64001} {\bibfield
   {journal} {\bibinfo  {journal} {Europhys. Lett.}\ }\textbf {\bibinfo
  {volume} {86}},\ \bibinfo {pages} {64001} (\bibinfo {year}
  {2009})}\BibitemShut {NoStop}%
\bibitem [{\citenamefont {Lauga}(2014)}]{lauga14}%
  \BibitemOpen
  \bibfield  {author} {\bibinfo {author} {\bibfnamefont {E.}~\bibnamefont
  {Lauga}},\ }\href {\doibase 10.1063/1.4891969} {\bibfield  {journal}
  {\bibinfo  {journal} {Phys. Fluids}\ }\textbf {\bibinfo {volume} {26}},\
  \bibinfo {pages} {081902} (\bibinfo {year} {2014})}\BibitemShut {NoStop}%
\bibitem [{\citenamefont {Elfring}\ and\ \citenamefont
  {Goyal}(2016)}]{elfring16}%
  \BibitemOpen
  \bibfield  {author} {\bibinfo {author} {\bibfnamefont {G.~J.}\ \bibnamefont
  {Elfring}}\ and\ \bibinfo {author} {\bibfnamefont {G.}~\bibnamefont
  {Goyal}},\ }\href {\doibase 10.1016/j.jnnfm.2016.04.005} {\bibfield
  {journal} {\bibinfo  {journal} {J. Non-Newtonian Fluid Mech.}\ }\textbf
  {\bibinfo {volume} {234}},\ \bibinfo {pages} {8 } (\bibinfo {year}
  {2016})}\BibitemShut {NoStop}%
\bibitem [{\citenamefont {Elfring}(2017)}]{elfring17}%
  \BibitemOpen
  \bibfield  {author} {\bibinfo {author} {\bibfnamefont {G.~J.}\ \bibnamefont
  {Elfring}},\ }\href {\doibase 10.1017/jfm.2017.632} {\bibfield  {journal}
  {\bibinfo  {journal} {J. Fluid Mech.}\ }\textbf {\bibinfo {volume} {829}},\
  \bibinfo {pages} {R3} (\bibinfo {year} {2017})}\BibitemShut {NoStop}%
\bibitem [{\citenamefont {Elfring}(2015)}]{elfring15}%
  \BibitemOpen
  \bibfield  {author} {\bibinfo {author} {\bibfnamefont {G.~J.}\ \bibnamefont
  {Elfring}},\ }\href {\doibase http://dx.doi.org/10.1063/1.4906993} {\bibfield
   {journal} {\bibinfo  {journal} {Phys. Fluids}\ }\textbf {\bibinfo {volume}
  {27}},\ \bibinfo {pages} {023101} (\bibinfo {year} {2015})}\BibitemShut
  {NoStop}%
\bibitem [{\citenamefont {Yariv}\ and\ \citenamefont
  {Michelin}(2015)}]{yariv15}%
  \BibitemOpen
  \bibfield  {author} {\bibinfo {author} {\bibfnamefont {E.}~\bibnamefont
  {Yariv}}\ and\ \bibinfo {author} {\bibfnamefont {S.}~\bibnamefont
  {Michelin}},\ }\href {\doibase 10.1017/jfm.2015.78} {\bibfield  {journal}
  {\bibinfo  {journal} {J. Fluid Mech.}\ }\textbf {\bibinfo {volume} {768}},\
  \bibinfo {pages} {R1} (\bibinfo {year} {2015})}\BibitemShut {NoStop}%
\bibitem [{\citenamefont {Patil}\ \emph {et~al.}(2006)\citenamefont {Patil},
  \citenamefont {Feng},\ and\ \citenamefont
  {Hatzikiriakos}}]{patil2006constitutive}%
  \BibitemOpen
  \bibfield  {author} {\bibinfo {author} {\bibfnamefont {P.~D.}\ \bibnamefont
  {Patil}}, \bibinfo {author} {\bibfnamefont {J.~J.}\ \bibnamefont {Feng}}, \
  and\ \bibinfo {author} {\bibfnamefont {S.~G.}\ \bibnamefont
  {Hatzikiriakos}},\ }\href {\doibase
  http://dx.doi.org/10.1016/j.jnnfm.2006.05.013} {\bibfield  {journal}
  {\bibinfo  {journal} {J. Non-Newton. Fluid Mech.}\ }\textbf {\bibinfo
  {volume} {139}},\ \bibinfo {pages} {44 } (\bibinfo {year}
  {2006})}\BibitemShut {NoStop}%
\end{thebibliography}%

\end{document}